\documentclass[aps, prd, amsmath, floats,floatfix, twocolumn, superscriptaddress, nofootinbib, showpacs]{revtex4}

\usepackage{amssymb}
\usepackage{amsmath}
\usepackage{verbatim}
\usepackage{mathrsfs}
\usepackage{amsfonts}
\usepackage{latexsym}
\usepackage{epsfig}
\usepackage{epstopdf}
\usepackage{color}
\usepackage{graphicx,subfigure}
\usepackage{units}

\newcommand{\re}{\mbox{Re}}

\begin{document}


\definecolor{orange}{rgb}{0.9,0.45,0} 

\newcommand{\argelia}[1]{\textcolor{red}{{\bf Argelia: #1}}}
\newcommand{\dario}[1]{\textcolor{red}{{\bf Dario: #1}}}
\newcommand{\juanc}[1]{\textcolor{green}{{\bf JC: #1}}}
\newcommand{\juan}[1]{\textcolor{cyan}{{\bf Juan B: #1}}}
\newcommand{\alberto}[1]{\textcolor{blue}{{\bf Alberto: #1}}}
\newcommand{\miguela}[1]{\textcolor{red}{{\bf Miguel: #1}}}
\newcommand{\mm}[1]{\textcolor{orange}{{\bf MM: #1}}}
\newcommand{\OS}[1]{\textcolor{blue}{{\bf Olivier: #1}}}

\long\def\symbolfootnote[#1]#2{\begingroup%
\def\thefootnote{\fnsymbol{footnote}}\footnote[#1]{#2}\endgroup}


\title{Self-gravitating black hole scalar wigs}

\author{Juan~Barranco} 
\affiliation{Departamento de F\'isica, Divisi\'on de Ciencias e Ingenier\'ias, Campus Le\'on, 
Universidad de Guanajuato, Le\'on 37150, M\'exico}

\author{Argelia~Bernal} 
\affiliation{Departamento de F\'isica, Divisi\'on de Ciencias e Ingenier\'ias, Campus Le\'on, 
Universidad de Guanajuato, Le\'on 37150, M\'exico}
\affiliation{\'Area Acad\'emica de Matem\'aticas y F\'isica, Universidad
Aut\'onoma del Estado de Hidalgo, Carretera Pachuca-Tulancingo Km. 4.5, C.P. 42184,
Pachuca, Hidalgo, M\'exico}

\author{Juan~Carlos~Degollado} 
\affiliation{Instituto de Ciencias F\'isicas, Universidad Nacional Aut\'onoma de M\'exico,
Apdo. Postal 48-3, 62251, Cuernavaca, Morelos, M\'exico}

\author{Alberto~Diez-Tejedor}
\affiliation{Departamento de F\'isica, Divisi\'on de Ciencias e Ingenier\'ias, Campus Le\'on, 
Universidad de Guanajuato, Le\'on 37150, M\'exico}

\author{Miguel~Megevand}
\affiliation{Instituto de F\'isica Enrique Gaviola, CONICET. Ciudad Universitaria, 5000 C\'ordoba, Argentina}

\author{Dar\'{\i}o~N\'u\~nez}
\affiliation{Instituto de Ciencias Nucleares, Universidad Nacional
  Aut\'onoma de M\'exico, Circuito Exterior C.U., A.P. 70-543,
  M\'exico D.F. 04510, M\'exico}

\author{Olivier~Sarbach}
\affiliation{Instituto de F\'{\i}sica y Matem\'aticas, Universidad
Michoacana de San Nicol\'as de Hidalgo, Edificio C-3, Ciudad
Universitaria, 58040 Morelia, Michoac\'an, M\'exico}


\date{\today}


\begin{abstract}  
It has long been known that no static, spherically symmetric, asymptotically flat Klein-Gordon scalar field configuration surrounding a nonrotating black hole can exist in general 
relativity. In a series of previous papers we proved that, at the effective level, this no-hair theorem can be circumvented by relaxing the staticity assumption: for appropriate 
model parameters there are quasi-bound scalar field configurations living on a fixed Schwarzschild background which, although not being strictly static, have a larger lifetime than 
the age of the universe. This situation arises when the mass of the scalar field distribution is much smaller than the black hole mass, and following the analogies with the hair in the literature 
we dubbed these long-lived field configurations {\it wigs}. Here we extend our previous work to include the gravitational backreaction produced by the scalar wigs. 
We derive new approximate solutions of the spherically symmetric Einstein-Klein-Gordon system which represent self-gravitating scalar wigs surrounding black holes. 
These configurations interpolate between boson star configurations and Schwarzschild black holes dressed with the long-lived scalar test field distributions discussed in previous 
papers. Nonlinear numerical evolutions of initial data sets extracted from our approximate solutions support the validity of our approach. 
Arbitrarily large lifetimes are still possible, although for the parameter space that we analyze in this paper they seem to decay faster than the quasi-bound states.
Finally, we speculate about the possibility that these configurations could 
describe the innermost regions of dark matter halos.
\end{abstract}




\maketitle


\section{Introduction}
\label{sec:introduction}

There is now compelling evidence for the existence of a fundamental 
scalar field in nature, given the recent discovery of a new particle at the LHC 
consistent with the Higgs boson in the standard model~\cite{Aad:2012tfa, Chatrchyan:2012ufa}. 
At a more phenomenological level, however, scalar fields with a deeper 
substructure have been well-known at least since the dawn of nuclear physics.
Nowadays scalar fields are widely used in particle physics, in string theory, 
and also in cosmology, representing an active area of research in gravitational 
physics~\cite{Classical}.

Along these lines, axion-like particles and other light scalar fields have been 
considered as possible candidates for  dark 
matter~\cite{Suarez:2013iw,Marsh:2015xka,Hui:2016ltb}. 
At cosmological scales, the highly populated, low-energy excitations of a 
scalar field could drive the expansion of the universe during the matter 
dominated era~\cite{Preskill:1982cy,Abbott:1982af,Dine:1982ah}. Possible 
mechanisms to excite these long-wavelength modes in a macroscopic way include a 
regime of thermal equilibrium with 
the standard model sector giving rise to the appearance of a cosmological Bose-Einstein 
condensate~\cite{UrenaLopez:2008zh,Lundgren:2010sp,Li:2013nal,Aguirre:2015mva}, 
or a nonthermal vacuum realignment of the axion potential in the early 
universe~\cite{Marsh:2014qoa, Diez-Tejedor:2017ivd}. Furthermore, it has been 
shown that the large scale structure that emerges as a consequence of the 
gravitational instability of the primordial perturbations in this matter 
component is consistent with observations as long as the mass of the scalar 
particle is not too light, $\mu \gtrsim 
10^{-22}\,\textrm{eV}$~\cite{Bozek:2014uqa,Schive:2015kza,Urena-Lopez:2015gur,
Corasaniti:2016epp}. The QCD axion is probably the best motivated and well-known 
realization of this 
scenario~\cite{Peccei:2006as,Kim:2008hd,Sikivie:2006ni,Kawasaki:2013ae}, 
although there have been many other proposals with a similar 
spirit~\cite{Sin:1992bg,Lee:1995af,Hu:2000ke,Matos:2000ss,Goodman:2000tg,
Marsh:2010wq,Evslin:2012fe} motivated to some extent by the absence of any 
direct or indirect positive signal of the existence of thermal WIMPs in 
astrophysical observations and/or ground-based detectors.  

At the scale of galaxies, the wave properties of an ultralight matter 
component could in addition alleviate some of the classic discrepancies of the 
standard cold dark matter 
model~\cite{Suarez:2013iw,Marsh:2015xka,Hui:2016ltb}.  On the one hand, the 
presence of light particles in the universe suppresses the mass power spectrum 
below a characteristic 
length scale. For the highly populated zero-mode of a dark matter boson with a 
mass of the order of $\mu\sim 10^{-22}\,\textrm{eV}$ 
this scale can be of astrophysical interest, reducing the number of low-mass halo objects from gravitational instability and alleviating in this way the ``missing satellites 
problem''~\cite{2010arXiv1009.4505B}. Moreover, using cosmological nonlinear codes 
for the study of structure formation, large concentrations of ultralight 
low-energy scalars have been identified to smooth the innermost regions of 
galactic halos through the formation of solitonic wave-like objects in their 
centers~\cite{Schive:2014dra,Schive:2014hza,Schwabe:2016rze,Veltmaat:2016rxo,
Du:2016aik}. These boson stars admit a description in terms of a {\it 
classical} field theory~\cite{Ruffini69,Liebling:2012fv}, and more interestingly can alleviate the 
so called ``cusp/core problem'' of the standard cosmological 
scenario~\cite{2010AdAst2010E...5D}. Although possible sources of tension 
between cosmological and galactic observations in this model have been recently 
identified in e.g. Refs.~\cite{Marsh:2015wka, Gonzales-Morales:2016mkl, Irsic:2017yje}, an 
ultralight scalar dark matter particle is today a hot research topic of modern 
cosmology. 

However, these solitonic coherent macroscopic scalar excitations in the galactic center are likely not alone since most galaxies are expected to host supermassive black 
holes. It has been known for some time that if a galactic dark matter halo is 
described in terms of a boson star, then such a configuration cannot live forever~\cite{Pena:1997cy}. The interaction of 
an ultralight scalar field with a black hole has been considered in some detail 
in Refs.~\cite{UrenaLopez:2002du, CruzOsorio:2010qs, UrenaLopez:2011fd, 
Guzman:2012jc, Burt:2011pv, Barranco:2012qs, Barranco:2013rua, Helfer:2016ljl}, 
and extremely long-lived scalar field configurations (referred to as {\em quasi-bound states} in the following) have been constructed in 
the test-field regime of the theory~\cite{Burt:2011pv, Barranco:2012qs, Barranco:2013rua}. 
Moreover, the distribution of a massive scalar field surrounding a black hole has been 
shown to give rise to very rich 
dynamics~\cite{Berti:2015itd, Okawa:2014nda, Herdeiro:2015waa, Cardoso:2014uka} and to field configurations that fall off so slowly to the point that they may co-exist with the black hole for cosmological times~\cite{Burt:2011pv, 
Barranco:2012qs}. These results encourage further studies of the properties of fundamental fields as a solid candidates to describe the dark matter, providing an alternative to the 
more usual particle-like description.

In a series of previous papers~\cite{Burt:2011pv,Barranco:2012qs,Barranco:2013rua} we provided a detailed analysis for the propagation of a massive Klein-Gordon field on a fixed, 
Schwarzschild background. Among other results, we showed that when the Compton wavelength of the field is much larger than the Schwarzschild radius of the black hole, 
configurations which persist in the vicinity of the horizon for cosmological times do exist~\cite{Burt:2011pv}, that such field distributions are generic~\cite{Barranco:2012qs}, 
and that their late time behavior from arbitrary initial configurations can be described using semi-analytic calculations without the need of performing a Cauchy 
evolution~\cite{Barranco:2013rua}. 

Clearly, there are several natural extensions of this work towards a more realistic model. One pressing goal is the inclusion of the 
gravitational backreaction of the scalar field that has been neglected in our previous calculations, implying that our results so far are only applicable to distributions whose 
mass is small enough such that the gravitational field is dominated by the one generated by the black hole. However, this is clearly not the case for dark matter halos 
surrounding the black hole at the galactic centers. A further goal is to take into account the angular momentum of the black hole, although the effects of the rotation are 
probably only important for an accurate description of the scalar field distribution in the region close to the event horizon. Some work in these directions has already been 
carried out by different groups.

Numerical calculations of the full nonlinear Einstein-Klein-Gordon system of equations for the evolution of massive and massless scalar fields around non-rotating and rotating black holes have been performed in Ref.~\cite{Okawa:2014nda}.
The authors report that some fraction of the initial scalar field is accreted by the black 
hole, leading to an increase in its mass, while another fraction remains in the 
vicinity of the event horizon with a distribution resembling that of quasi-bound 
states in the test-field approximation. In Ref.~\cite{Sanchis-Gual:2014ewa}, long term evolutions of spherically symmetric scalar field configurations around 
black holes were also studied by numerical means.
Using Gaussian packets as initial data for 
the scalar field, the evolutions yield configurations describing nonrotating 
black holes surrounded by scalar clouds. It was found that, for initial 
configurations consisting of a black hole with a rich scalar environment, after 
a highly dynamical stage the evolution 
settles down to states which are consistent with the quasi-bound states. 
In both works~\cite{Okawa:2014nda,Sanchis-Gual:2014ewa} it has been found that the scalar field oscillates with a combination of well-defined frequencies, 
and exhibits a characteristic beating pattern which is due to the combination of two frequencies lying close to each other, as predicted by previous calculations in the test-field approximation~\cite{Witek:2012tr}. 
Further numerical work~\cite{Herdeiro:2014goa,Herdeiro:2015gia} has revealed the existence of stationary solutions of the full nonlinear Einstein-Klein-Gordon equations 
branching off the Kerr solution (dubbed Kerr black holes with scalar hair), provided the rotational parameter of the underlying Kerr black hole assumes some special values. 
This shows that infinitely long-lived configurations of scalar fields may survive in the vicinity of certain spinning black holes.

The present manuscript provides new results regarding the description of self-gravitating scalar field configurations in the presence of nonrotating black holes, and it extends our 
previous work in Refs.~\cite{Burt:2011pv, Barranco:2012qs, Barranco:2013rua} to the regime in which the gravitational backreaction of the scalar field becomes relevant.
In particular, we present a new method for constructing initial data sets which are devised to represent a time slice of a scalar field distribution surrounding a black hole and surviving the presence of the horizon for extremely long periods of time.
Similar states have been previously identified in numerical simulations, but contrary to our construction, these simulations were performed starting from either rather generic 
initial configurations or from initial data motivated by results from the test-field limit, where large amounts of the field content are lost in the initial stages of the 
evolution. Furthermore, we also present for the first time an approximate 
semi-analytic description of these self-gravitating objects, dubbed 
{\it black hole scalar wigs} in this paper, that can be used to explore their characteristic 
frequencies of oscillation and decaying time scales without the need of 
performing a time evolution.

In order to achieve this, we introduce an approximation method for solving the Einstein-Klein-Gordon system in the presence of black holes.
This technique consists in proposing a coherent state ansatz similar to that of a static boson star, with the important difference that the standard regularity conditions at the 
origin are replaced by an appropriate ``horizon boundary condition", where purely outgoing boundary conditions are imposed for the scalar field. The configurations constructed 
in this way do not yield exact solutions of the Einstein-Klein-Gordon equations; nevertheless, and as we show, they provide a model for black hole scalar wigs which is accurate 
for large time spans. Additionally, our configurations can be used to construct {\it exact} initial data sets. Based on a numerical evolution of the full Einstein-Klein-Gordon 
system starting with such sets we prove the viability of our approximation method.

The remaining part of this work is organized as follows. In Sec.~\ref{Sec:EoM} we present the Einstein-Klein-Gordon system as well as a simple method for constructing spherically 
symmetric initial data sets for this problem. Next, in Sec.~\ref{Sec:Review} we review well-known solutions describing static boson stars or quasi-bound states living on a 
Schwarzschild background and mention some of their most important properties that are relevant for the investigation that follows. In Sec.~\ref{Sec:Results} we present the main result of this paper, namely our approximation method for computing long-lived, self-gravitating scalar field configurations surrounding black holes. This method leads to a nonlinear radial eigenvalue problem which is solved numerically. By construction, these new configurations interpolate between the static boson stars and the quasi-bound states on the fixed Schwarzschild background, and the verification of these limits from the numerical data is also shown in Sec.~\ref{Sec:Results}. Next, in Sec.~\ref{Sec:Numerical} we perform numerical evolutions of the spherically symmetric Einstein-Klein-Gordon equations without further approximations, starting with initial data corresponding to a fixed time slice of the solutions constructed in Sec.~\ref{Sec:Results}. We find that the solutions computed from the numerical time evolution slowly decay in time and oscillate at rates which are in excellent agreement with the approximate solutions, validating our approach.
Finally, conclusions and a discussion about the lifetime of some of the black hole scalar wigs models discussed in this article are given in Sec.~\ref{Sec:Discussion}.

Throughout this paper we work with Planck units, in which the gravitational constant, the speed of light and Planck's constant are set to one, such that all quantities are 
dimensionless. Additionally, we use the signature convention $(-,+,+,+)$ for the spacetime metric.

\section{The spherical Einstein-Klein-Gordon problem}
\label{Sec:EoM}

In this section, we briefly review the equations of motion describing a spherically symmetric, self-gravitating scalar field configuration and introduce the choice of variables and 
gauge conditions we find useful for our investigation. The main result of this section is the presentation of the Einstein-Klein-Gordon system in the form of Eqs.~(\ref{eq:ESS}) and~(\ref{eq:dyn.scalar}). As a by-product, we also 
obtain a simple method of constructing spherically symmetric initial data sets.

Without loss of generality we can parametrize the spherically symmetric spacetime line-element in terms of Arnowitt-Deser-Misner (ADM) variables,
\begin{equation}
ds^2 = -\alpha^2 dt^2 + \gamma^2(dr + \beta dt)^2 
 + r^2 d\Omega^2.
 \label{Eq:gADM}
\end{equation}
Here $\alpha$ denotes the lapse function, $\beta$ the radial component of the shift vector, $\gamma^2$ the radial-radial component of the three-metric, and 
$d\Omega^2=d\vartheta^2 + \sin^2\vartheta\, d\varphi^2$ the standard line-element on the unit two-sphere. 
Due to the spherical symmetry and choice of coordinates, these quantities depend only on time $t$ and the areal radius $r$, but not on the angular 
variables $\vartheta$ and $\varphi$.\footnote{In the standard ADM formalism the lapse function and the shift vector are usually associated with the gauge 
degrees of freedom corresponding to the choice of the spacetime coordinates. In our case, the spatial coordinates are fixed by choosing the areal radius $r$ and the spherical 
coordinates $\vartheta$ and $\varphi$. Therefore, we do not have the freedom of choosing both $\alpha$ and $\beta$ arbitrarily anymore; however, we can still choose a combination 
of these two quantities, as we will do later.}

Although the parametrization of the metric tensor in terms of the variables $(\alpha,\beta,\gamma)$ is useful for many calculations, in this paper we find it more convenient to work with the slightly different set of quantities $(\alpha,\nu,m)$. Here, as before, the function $\alpha$ denotes the lapse, which determines the future-directed unit normal covector $n_{\mu} = -\alpha\nabla_{\mu}t$ to the slices of constant time $t = \textrm{const.}$, 
the function $\nu$ refers to the quantity
\begin{equation}
\nu := - g^{\mu\nu}(\nabla_{\mu}r) n_{\nu},
\end{equation}
which is slice-dependent as well, whereas $m$ is the Misner-Sharp mass function~\cite{cMdS64} which is invariantly defined in terms of the areal radius $r$ and its differential $\nabla_{\mu}r$ as follows:
\begin{equation}
1 - \frac{2m}{r} := g^{\mu\nu}(\nabla_{\mu}r)(\nabla_{\nu}r).
\label{Eq:MSMass}
\end{equation}
The relation between the ADM variables $(\alpha,\beta,\gamma)$ and the new set of quantities $(\alpha,\nu,m)$ that we use in this paper is given by
\begin{equation}
\beta = \alpha\nu,\quad
\gamma = \left( 1 - \frac{2m}{r} + \nu^2 \right)^{-1/2}.
\label{Eq:gamma}
\end{equation}

The functions $(\alpha,\nu,m)$ are related to the matter distribution through Einstein's field equations, $G_{\mu\nu}=\kappa T_{\mu\nu}$, which imply
\begin{subequations}\label{eq:ESS}
\begin{eqnarray}
D_0(2m) &=& \kappa r^2
\left( \nu S - \gamma^{-1} j \right),
\label{Eq:D0m}\\
2m' &=& \kappa r^2\left( \rho - \gamma\nu j \right),
\label{Eq:mprime}\\
(\log\alpha)' &=& \gamma^2 \left( -D_0\nu  + \frac{m}{r^2} 
 + \frac{\kappa r}{2}S \right).
\label{Eq:D0nu}
\end{eqnarray}
\end{subequations}
Here $\kappa= 8\pi$ is the gravitational coupling constant (remember that we work in Planck units), $D_0 := n^{\mu}\partial_{\mu} = \alpha^{-1}\partial_t - \nu\partial_r$ 
is the derivative in the normal direction to the hypersurfaces of constant time, and the prime denotes the derivative with respect to $r$. The source terms $\rho$, $j$ and $S$ are defined as the following contractions of the stress energy-momentum tensor $T_{\mu\nu}$ with the normal vector $n^\mu$ and the unit radial vector 
field $w^\mu = \gamma^{-1}\delta^\mu{}_r$ orthogonal to it:
\begin{equation}
\rho:=n^\mu n^\nu T_{\mu\nu},\quad
j := -n^\mu w^\nu T_{\mu\nu},\quad
S := w^\mu w^\nu T_{\mu\nu}.
\end{equation}
So far these expressions are general and are applicable to any matter fields in spherical symmetry.

However, in this paper we will restrict our attention to the macroscopic, coherent, solitonic excitations of a canonical complex scalar field $\Phi(t,r)$ with an internal $U(1)$ 
global symmetry and, for simplicity, no self-interactions. For this matter component the energy-momentum tensor can be expressed in the form
\begin{equation}
T_{\mu\nu} = \re\left( \nabla_\mu\Phi^*\nabla_\nu\Phi \right)
 - \frac{1}{2} g_{\mu\nu}\left( \nabla_\alpha\Phi^*\nabla^\alpha\Phi + \mu^2\Phi^*\Phi \right),
\end{equation}
such that
\begin{subequations}\label{eq:source}
\begin{eqnarray}
\rho &=& \frac{1}{2}\left( |\Pi|^2 + |\chi|^2 +\mu^2|\Phi|^2\right),
\label{eq:den_flux}\\
S &=& \frac{1}{2}\left( |\Pi|^2 + |\chi|^2 -\mu^2|\Phi|^2\right),\\
j &=& -\re( \Pi^*\chi ). 
\end{eqnarray}
\end{subequations}
Here $\Phi^*$ denotes the complex conjugate of $\Phi$, $\Pi := D_0\Phi$ is the conjugate momentum associated with the scalar field, $\chi := \gamma^{-1}\Phi'$, and $\mu$
is the mass of the corresponding scalar particle.

The Klein-Gordon equation $(\nabla^\nu\nabla_\nu - \mu^2)\Phi = 0$ for the dynamical evolution of the scalar field can be cast into the following form:
\begin{subequations}\label{eq:dyn.scalar}
\begin{eqnarray}
D_0\Phi &=& \Pi,
\label{Eq:Phi}\\
D_0\Pi &=& \frac{1}{\alpha\gamma r^2}\left( r^2\frac{\alpha}{\gamma}\Phi' \right)'
 - K\Pi - \mu^2\Phi.
\label{Eq:Pi}
\end{eqnarray}
\end{subequations}
Herein, $K$ is the trace of the extrinsic curvature associated with the constant time slices, which can be computed from the functions $\nu$, $\gamma$ and $j = -\re(\Pi^*\chi)$, see below for details. 
As previously discussed in this section, $\nu$ is a gauge parameter which can be freely 
specified. The lapse function $\alpha$ can then be obtained by integrating Eq.~(\ref{Eq:D0nu}), the radial-radial component of the three-metric $\gamma^2$ from Eq.~(\ref{Eq:gamma}), and the mass function $m$ can be computed from Eq.~(\ref{Eq:D0m}), from Eq.~(\ref{Eq:mprime}), or from a combination of the two.

We end this section with the following important remark. By solving the single equation~(\ref{Eq:mprime}), we may determine a spherically symmetric initial data set satisfying the 
constraint equations of the Einstein-Klein-Gordon system. To demonstrate this result, we first note that the components of the extrinsic curvature associated with the 
hypersurfaces of constant time $t$ are
\begin{subequations}
\begin{eqnarray}
k^r{}_r &=& \frac{1}{\gamma} D_0\gamma - \frac{1}{\alpha}(\alpha\nu)',
\label{Eq:krr}\\
k^r{}_B &=& 0,
\label{Eq:krB}\\
k^A{}_B &=& -\frac{\nu}{r}\delta^A{}_B,
\label{Eq:kAB}
\end{eqnarray}
\end{subequations}
where $A,B = \vartheta,\varphi$. Next, using Eqs.~(\ref{Eq:D0m}) and~(\ref{Eq:D0nu}) we can rewrite the radial-radial component as
\begin{equation}
k^r{}_r = -\nu' - \frac{\kappa r}{2}\gamma j.
\label{Eq:krrBis}
\end{equation}
Now let $m(r)$ and $\nu(r)$ be any two smooth functions satisfying Eq.~(\ref{Eq:mprime}) with $\gamma = (1 - 2m/r + \nu^2)^{-1/2}$. Then, we claim that the three-metric 
$\gamma(r)^2 dr^2 + r^2d\Omega$ and extrinsic curvature whose components are given by the expressions in Eqs.~(\ref{Eq:krrBis}), (\ref{Eq:krB}) and~(\ref{Eq:kAB}) satisfy the Hamiltonian and momentum constraint equations. To check the claim, we first notice that momentum constraint in spherical symmetry yields the single equation
\begin{equation}
k^r{}_r - \frac{1}{2}(r k^A{}_A)' = -\frac{\kappa r}{2}\gamma j,
\end{equation}
which is trivially satisfied by virtue of Eqs.~(\ref{Eq:kAB}) and~(\ref{Eq:krrBis}). Next, the Hamiltonian constraint is a consequence of the momentum constraint and 
Eq.~(\ref{Eq:mprime}), and the claim is then proven.

Therefore, given any initial scalar field configuration $(\Phi(r),\Pi(r))$ and a choice for the gauge function $\nu(r)$ one can generate an initial data set for the spherically symmetric Einstein-Klein-Gordon system by solving the single equation~(\ref{Eq:mprime}) for the mass function $m(r)$. Subsequently, this data may be evolved in time by means of Eqs.~(\ref{eq:ESS}) and~(\ref{eq:dyn.scalar}) and a suitable gauge condition for $\nu$. We will use this procedure to construct appropriate initial data sets representing a moment of time of self-gravitating scalar wigs surrounding black holes in Sec.~\ref{Sec:Results}. These data sets will be evolved numerically in Sec.~\ref{Sec:Numerical} using an independent method.

\section{Brief review of static boson stars and of quasi-bound states surrounding black holes}
\label{Sec:Review}

In this section, we provide a brief overview of two different well-known applications of the spherically symmetric Einstein-Klein-Gordon system, Eqs.~(\ref{eq:ESS}) and~(\ref{eq:dyn.scalar}) above. In both of them the scalar field describes a coherent state of the form
\begin{equation}
\Phi(t,r) = e^{st}\psi(r),
\label{Eq:Coherent}
\end{equation}
with $s = \sigma + i\omega$ a complex constant and $\psi(r)$ a complex-valued function depending on the areal radius coordinate $r$ only. Here $\sigma\le 0$ denotes the 
decay rate of the state, while $\omega$ describes the frequency of its oscillatory behavior. Alternatively, one might think of $- i s = \omega - i\sigma$ as a complex 
frequency with positive imaginary part describing the decay rate.  
As we discuss next, the ansatz in Eq.~(\ref{Eq:Coherent}) is appropriate to describe both static boson stars and quasi-bound scalar field configurations surrounding a Schwarzschild black hole in the test field approximation. Although these solutions have been extensively studied in the literature, we review their most important properties in the next two subsections since they will be essential for 
the motivation of the more general construction for the self-gravitating scalar soliton wigs of Sec.~\ref{Sec:Results}.

\subsection{Static boson stars}
\label{Sec:boson stars}

Static boson stars are described in terms of globally regular, localized solutions to the spherically symmetric Einstein-Klein-Gordon system, where the field describes a coherent excitation of the form in Eq.~(\ref{Eq:Coherent}) with
$s = i\omega$ purely imaginary and $\psi(r)$ real-valued. In the gauge $\nu = 0$, the system in Eqs.~(\ref{Eq:mprime}), (\ref{Eq:D0nu}) and (\ref{eq:dyn.scalar}) reduces to
\begin{subequations}\label{Eq:bosonstars}
\begin{eqnarray}
 m' = \frac{\kappa r^2}{4\gamma^2}
\left[ \psi'^2 + \gamma^2\left(\mu^2+\frac{\omega^2}{\alpha^2}\right)\psi^2 \right],
\label{Eq:bosonstars.2} \\
 \frac{\alpha'}{\alpha} = \gamma^2\frac{m}{r^2} + \frac{\kappa r}{4}\left[ \psi'^2
 - \gamma^2\left(\mu^2- \frac{\omega^2}{\alpha^2}\right)\psi^2\right],
\label{Eq:bosonstars.3} \\
 \frac{1}{\alpha\gamma r^2}\left(r^2\frac{\alpha}{\gamma}\psi'\right)' 
 - \left(\mu^2 - \frac{\omega^2}{\alpha^2}\right)\psi=0,
\label{Eq:bosonstars.1}
\end{eqnarray}
\end{subequations}
where $\gamma = (1 - 2m/r)^{-1/2}$. 

Demanding regularity at the origin
\begin{subequations}\label{Eqs:BSOrigin}
\begin{eqnarray}
m(r=0) &=& 0,\\
\alpha(r=0) &=& \alpha_c,\\
\psi(r=0) &=& \psi_c,\\
 \psi'(r=0) &=& 0,
\end{eqnarray}
\end{subequations}
and asymptotic flatness at spacial infinity, $\psi(r\to \infty) = 0$, one obtains a nonlinear eigenvalue problem for the frequency $\omega$. Here $\psi_c$ and $\alpha_c$ are two 
free positive arbitrary constants. Note however that the system in Eqs.~(\ref{Eq:bosonstars}) is invariant under the rescaling $(\alpha,\omega)\mapsto \lambda(\alpha,\omega)$ with some positive constant parameter $\lambda$, and hence the value of the lapse at the origin $\alpha_c$ is not really physical, although one usually chooses this function in such a way that $\alpha(r\to\infty) = 1$. Solving this problem gives rise to a discrete family of frequencies
$\omega_n(\psi_c)$ for each central value of the scalar field $\psi_c$, where
$n = 1, 2, 3, \ldots$ labels the different possible solutions. More details about these states are given shortly.

Since the integration of the system is done numerically, in practice one can find the eigenfrequencies $\omega_n(\psi_c)$ by means of a shooting algorithm. To proceed, one integrates 
Eqs.~(\ref{Eq:bosonstars}) outwards, starting from the initial conditions in Eq.~(\ref{Eqs:BSOrigin}), setting $\alpha_c = 1$ for convenience, and fine-tunes the value of the frequency $\omega$ to match the asymptotically flat solution $\psi(r\to\infty)\sim \exp[-\sqrt{\mu^2-\omega^2/\alpha^2}\,r]$ at a large but finite value of the radial coordinate $r_{\textrm{max}}$,
\begin{subequations}\label{Eqs:BSAsymptotic2}
\begin{eqnarray}
 \psi(r_{\textrm{max}}) &=& \psi_m,\\
 \psi'(r_{\textrm{max}}) &=& -\sqrt{\mu^2 - \frac{\omega^2}{\alpha^2}}\,\psi_m,
\end{eqnarray}
\end{subequations}
increasing the value of $r_{\textrm{max}}$ until the shooting parameter converges numerically. We choose the solution as the one which satisfies the outer boundary conditions for 
some final $r_{\textrm{max}}$ within a given tolerance.
The ``correct'' values of $\alpha_c$ and $ \omega$ can then be restored a posteriori by making use of the rescaling $(\alpha,\omega)\mapsto \lambda(\alpha,\omega)$ 
of the Einstein-Klein-Gordon system, with the value of $\lambda$ computed such that
 $\alpha(r_{\textrm{max}}) \gamma(r_{\textrm{max}})=1$ (as occurs for the Schwarzschild solution) at the outermost point of the spatial numerical grid.
We will follow a similar procedure later in Sec.~\ref{Sec:Results} when constructing self-gravitating scalar field configurations in the presence of black holes.
 
Notice that in order for $\psi(r)$ to decay at infinity, one needs $|\omega| < \mu$, which puts an upper bound on the frequencies $\omega_n(\psi_c)$. The solution corresponding to its lowest possible magnitude, $n=1$, describes the ground state, for which the function $\psi(r)$ has no zeros. In contrast to this, the function $\psi(r)$ for the excited states with $n >1$ have $n-1$ nodes. However, only the ground state is stable under small 
perturbations~\cite{Lee89}, and for that reason only this mode can be relevant for late time phenomena. 

\begin{figure}
\includegraphics[angle=0,width=0.48\textwidth,clip]{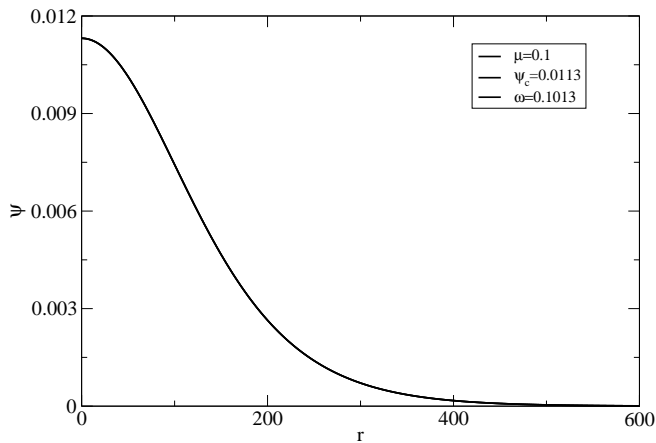}
\caption{The wave function $\psi(r)$ of a static boson star in the
ground state configuration, with $\mu=0.1$, $\psi_c = 0.0113$ and $\omega = 0.1013$.
}
\label{Fig:BSGroundState}
\end{figure}

For a given mass of the scalar particle, $\mu$, the field amplitude at the origin can be varied continuously from zero to $\psi_c^{\textrm{crit}}\sim 0.26$, increasing
in this way the mass $m(r\to\infty)$ of the self-gravitating object.
Configurations with an amplitude larger than $\psi_c^{\textrm{crit}}$ are also unstable against small perturbations~\cite{Gleiser:1988rq}, and thus they are not of interest for our purposes in this paper. A plot showing the typical profile of the function $\psi(r)$ for the ground state is exhibited in Fig.~\ref{Fig:BSGroundState}. 
In this case we have fixed $\mu=0.1$, although this profile can be easily rescaled to an arbitrary value of the scalar field mass using the invariance of the 
system in Eqs.~(\ref{Eq:bosonstars}) under the transformation
\begin{equation}\label{eq.sym1}
 \mu\to a\mu,\quad \omega\to a\omega,\quad r\to a^{-1}r, \quad m\to a^{-1}m,
\end{equation}
with $\psi$, $\alpha$ and $\gamma$ unchanged. In order to make contact with an ultralight axion dark matter component one would choose $a=8.2\times 10^{-51}\mu[10^{-22}\,\textrm{eV}]/\mu$, where here and in the following $\mu$ and $\mu[10^{-22}\,\textrm{eV}]$ refer to the scalar field mass in Planck units and units of $10^{-22}\,\textrm{eV}$, respectively. Likewise, the characteristic astrophysical mass and length scales can be obtained from the relations $m[10^8\,M_{\odot}]=1.3\times 10^{4}(m\mu)/\mu[10^{-22}\,\textrm{eV}]$ and 
$r[\textrm{pc}]=6.4\times 10^{-2}(\mu r)/\mu[10^{-22}\,\textrm{eV}]$, where $m$ and $r$ refer to the mass and radius in Planck units.
Further properties of the boson star solutions and their interpretation in terms of conserved quantities are discussed in Ref.~\cite{Liebling:2012fv}.

\subsection{Quasi-bound states on a Schwarzschild background}
\label{Sec:Quasi-bound states}

Like the case discussed in the previous subsection, quasi-bound states living on a Schwarzschild black hole background are obtained from the coherent state ansatz in Eq.~(\ref{Eq:Coherent}). However, contrary to the case of boson stars, we neglect now the self-gravity of the scalar field, which is equivalent to setting $\kappa = 0$ in Einstein's field 
equations. Under these assumptions, integration of Eqs.~(\ref{eq:ESS}) gives (up to a multiplicative constant which can be reabsorbed in the definition of $t$)
\begin{equation}
\alpha^2(r) = \frac{1}{\gamma^2(r)} = 1 - \frac{2m}{r} + \nu(r)^2,\quad
m = M_{\textrm{BH}},
 \label{eq:alpha_qbs}
\end{equation}
with $M_{\textrm{BH}}$ the (constant) mass of the black hole. The gauge choice $\nu = 0$ leads to the Schwarzschild metric in standard Schwarzschild coordinates, which are 
regular in the exterior region $2m < r < \infty$ but singular at the horizon, $r = 2m$. The alternative choice $\nu(r) = 2m/r(1 + 2m/r)^{-1/2}$, which we will use later, 
leads to ingoing Eddington-Finkelstein (iEF) coordinates, which are regular for all $0 < r < \infty$ including the future horizon.\footnote{In what follows, and in order to 
distinguish them from Schwarzschild coordinates, these coordinates will be denoted by $(t_{\textrm{EF}},r)$ and referred to as ``ingoing Eddington-Finkelstein coordinates" 
even though this name usually refers to the coordinates $(t_{\textrm{EF}} + r,r)$ in the literature~\cite{Wald84}.}

Introducing the ansatz in Eq.~(\ref{Eq:Coherent}) into the Klein-Gordon equation~(\ref{eq:dyn.scalar}), we obtain, in the gauge $\nu = 0$,
\begin{equation}
\frac{1}{r^2}\left(r^2\alpha^2\psi'\right)' - \left(\mu^2+\frac{s^2}{\alpha^2}\right)\psi=0 .
\label{eq:qbs}
\end{equation}
Note that the differential equation for the wave-function $\psi(r)$ has precisely the same form as the one in the boson star case, Eq.~(\ref{Eq:bosonstars.1}), except that 
now $s = \sigma + i\omega$ has in general a non-trivial negative real part $\sigma < 0$, and the metric coefficients $\alpha^2(r)$ and $\gamma^2(r)$ correspond to those of the Schwarzschild black hole, see Eq.~(\ref{eq:alpha_qbs}), instead of being determined by the self-gravity of the scalar excitation. 
Solutions to Eq.~(\ref{eq:qbs}) with a purely real frequency $-i s = \omega$ which decay at spatial infinity still exist; however, they are not regular on the future (nor the past) horizon, and in this sense they do not correspond to physical situations~\cite{Burt:2011pv}. 
As mentioned previously, the negative real part $\sigma < 0$ is necessary for the solution to decay in time and in this way evading the hypothesis of the no-hair theorems.

As we have discussed in detail in Ref.~\cite{Barranco:2013rua}, the quasi-bound states are obtained by demanding purely outgoing boundary conditions on the black hole horizon, and
can be expressed in terms of the confluent Heun function,
\begin{equation}
\psi(r) = z^{2M_{\textrm{BH}} s} e^{-\Omega(s)z}
\mbox{HeunC}\left( 2\Omega(s),4M_{\textrm{BH}} s,0,\delta,-\delta,-z \right),
\label{Eq:psiCloseBH}
\end{equation}
where $z := r/(2M_{\textrm{BH}}) - 1$, $\Omega(s) = 2M_{\textrm{BH}}\sqrt{\mu^2 + s^2}$, $\delta = \Omega^2(s) + (2M_{\textrm{BH}} s)^2$, and $\mbox{HeunC}$ is the confluent Heun 
function as defined in {\tt MAPLE}. By definition $\mbox{HeunC}$ is regular 
close to the origin and equal to one at $z=0$, so the expression in 
Eq.~(\ref{Eq:psiCloseBH}) provides 
a closed-form representation of the scalar field in the vicinity of the black hole horizon. 
(Note however that since $\textrm{Re}(s)<0$, the term $z^{2M_{\textrm{BH}}s}$ 
in Eq.~(\ref{Eq:psiCloseBH}) is not regular at $z=0$, so these scalar field 
configurations are still singular on $t = \textrm{const.}$ hypersurfaces. We will return to this point in a moment.) For certain discrete values of $s$, the function defined on the right-hand side of Eq.~(\ref{Eq:psiCloseBH}) decays to zero as $z\to \infty$, and these give rise to the quasi-bound states. 

Detailed studies of quasi-bound states in Schwarzschild and Kerr black holes have been given in several papers using different techniques, see for instance Refs.~\cite{Zouros:1979iw,Ternov:1978gq,Detweiler1980,Dolan:2007mj,Rosa:2011my,Dolan:2012yt}. Perhaps the most accurate way to  find the quasi-bound states is via the continued fraction method first employed by Leaver~\cite{Leaver:1985ax} in the context of the quasi-normal modes, and 
later used by Dolan~\cite{Dolan:2007mj} to find the frequencies of the 
quasi-bound sates, see also~\cite{Konoplya:2004wg,Konoplya:2011qq} and 
references therein. Furthermore, analytic expressions for the discrete frequencies, valid for small values of the dimensionless parameter $\varepsilon:= M_{\textrm{BH}}\mu$, were obtained long ago by Ternov {\it et al.}~\cite{Ternov:1978gq} and Detweiler~\cite{Detweiler1980}, and for $\ell=0$ they yield
\begin{equation}\label{eq.limitepsionsmall}
\sigma = -16\mu\varepsilon^5,\quad
\omega_n = \pm \mu\sqrt{1 - \frac{\varepsilon^2}{n^2}},
\end{equation}
where $n = 1,2,3,\ldots$. Although these expressions were only derived for the case with a nonvanishing angular momentum number (after decomposing the field in spherical harmonics), a direct comparison of the frequencies with those obtained from the continued fraction method also yields a very good agreement for the case with $\ell=0$~\cite{Barranco:2012qs}. 

These states can be interpreted by reformulating Eq.~(\ref{eq:qbs}) as a 
time-independent Schr\"odinger equation for the field $r\psi$, whose effective potential has a  well, see for instance Refs.~\cite{Burt:2011pv,Barranco:2012qs} for further details.
For the case of interest in this paper where $\ell =0$, the potential only develops a well if the product of the black hole and the scalar field masses satisfies the inequality 
$M_{\textrm{BH}}\mu<1/4$, i.e. $M_{\textrm{BH}}[10^8\,M_{\odot}]\mu[10^{-22}\,\textrm{eV}]\lesssim3.3\times 10^3$ in physical units,
so this is a necessary condition for the existence of quasi-bound states if no angular momentum is present in the field configuration~\cite{Burt:2011pv}.
The decay is then due to tunneling of the field through the potential barrier towards the horizon. Although they do probably not form a basis for the solution space, we have shown in~\cite{Barranco:2013rua} that the late time behavior of the scalar field can be 
accurately described by a linear combination of quasi-bound states, where the coefficients can be computed from the initial data alone by taking appropriate integrals.

\begin{figure}[]
\includegraphics[angle=0,width=0.48\textwidth,clip]{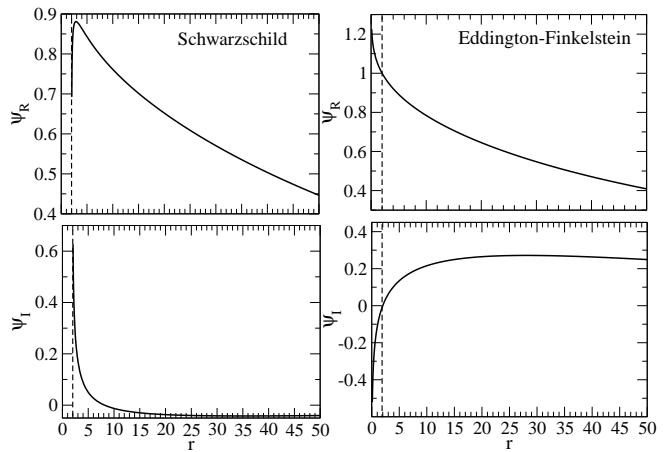}
\caption{{\it Left column:} The real and imaginary parts of the wave function $\psi(r)$ 
describing a quasi-bound state in the ground state configuration, with $M_{\textrm{BH}}=1$, $\mu=0.1$ and $s= -0.0000153092+0.09945i$, in Schwarzschild coordinates.
{\it Right column:} Same as in left column but in iEF coordinates.
Note that the use of iEF coordinates moves the divergence of the wave function from the horizon to the spacetime singularity.
}\label{fig:quasibound}
\end{figure}

The profile $\psi(r)$ of the ground state for the case in which $M_{\textrm{BH}}=1$ and $\mu=0.1$ is plotted in Fig.~\ref{fig:quasibound}. It is instructive to compare this plot with the one showing a typical profile of a static boson star in Fig.~\ref{Fig:BSGroundState}. In particular, note there are both real and imaginary contributions to the wave function when the black hole is present. There is again a scaling symmetry similar to that in Eq.~(\ref{eq.sym1}),
\begin{equation}\label{eq.sym2}
 \mu\to a\mu,\quad s\to as,\quad r\to a^{-1}r, \quad M_{\textrm{BH}}\to a^{-1}M_{\textrm{BH}},
\end{equation}
with $\psi$ unchanged, that makes it possible to scale the quasi-bound sates to 
an arbitrary value of the scalar field mass. 
Note that the product $M_{\textrm{BH}}\mu$ is invariant under this symmetry 
transformation, so we can use the solution in Fig.~\ref{fig:quasibound} 
to reach any value of $M_{\textrm{BH}}$ and $\mu$ in the combination $M_{\textrm{BH}}\mu=0.1$. 

For a typical scenario involving an ultralight scalar field one obtains, using the same conversion factor $a=8.2\times 10^{-51}\mu[10^{-22}\,\textrm{eV}]/\mu$ as in the case of the static bosons stars,
\begin{equation}\label{eq.timephys}
 t_{1/2}[\textrm{years}] \sim \frac{0.2}{\mu[10^{-22}\,\textrm{eV}]}
\left(\frac{\mu}{|\sigma|}\right)
\end{equation}
for the half-lifetime of a quasi-bound state in physical units, where $\mu$ and $\sigma$ are expressed in Planck units.
For the case of a ground state with no angular momentum, $\ell=0$, and as long as the inequality $M_{\textrm{BH}}[10^8\,M_{\odot}]\mu[10^{-22}\,\textrm{eV}]\ll 1.3\times 10^4$ 
between the black hole and scalar field masses is satisfied, 
the expression for the decay rate in Eq.~(\ref{eq.limitepsionsmall}) provides a time scale that is of cosmological interest, namely
\begin{equation}\label{eq.t1/2test}
 t_{1/2}[\textrm{years}] \sim \frac{5.6\times 10^{18}}{M_{\textrm{BH}}^5[10^8\,M_{\odot}]\mu^6[10^{-22}\,\textrm{eV}]},
\end{equation}
as has been previously noticed in Refs.~\cite{Burt:2011pv, Barranco:2012qs, Barranco:2013rua}. However, we emphasize that this expression does not take into account the self-gravity of the boson particles that would constitute the dark matter halo. The correct half-lieftime of these objects which takes into account the gravitational backreaction will be presented in Sec.~\ref{Sec:Discussion}.

We end this section with the following important remark concerning the energy associated with the quasi-bound states. The energy of the scalar field contained in a constant $t$ 
surface is given by (see Appendix~\ref{App:ConservationLaws})
\begin{equation}
E(t) = 4\pi\int\limits_{2m}^\infty \rho r^2 dr,
\end{equation}
where the energy density for the coherent state ansatz in Eq.~(\ref{Eq:Coherent}) turns out to be
\begin{equation}\label{eq.energydensity}
\rho = \frac{1}{2} e^{2\sigma t}\left[ \alpha^2|\psi'|^2 
 + \left(  \mu^2 + \frac{|s|^2}{\alpha^2} \right) |\psi|^2 \right].
\end{equation}
Close to the horizon, the term $\alpha^2|\psi'|^2$ diverges as $z^{4M_{\textrm{BH}} \sigma - 1}$, and since $\sigma < 0$, this implies that $E(t)$ is infinite. Therefore, the test 
scalar field has infinite energy on Schwarzschild $t = \textrm{const.}$ slices~\cite{Dolan:2007mj,Burt:2011pv}. In the self-gravitating case of Sec.~\ref{Sec:Results}, such an infinite energy will be clearly problematic, since it implies an infinite change in 
the mass function, see Appendix~\ref{App:ConservationLaws}.

On the other hand, when measuring the energy of the scalar field contained in a spacelike slice that penetrates the future horizon, one obtains finite expressions. 
For instance, choosing iEF coordinates, where the time coordinate $t$ is replaced with
\begin{equation}
t_{\textrm{EF}} = t + 2M_{\textrm{BH}}\ln\left(r/2M_{\textrm{BH}}-1\right)
\label{Eq:tEFDef}
\end{equation}
the ansatz in Eq.~(\ref{Eq:Coherent}) takes the form
\begin{equation}
\Phi(t,r) = e^{s t_{\textrm{EF}}} \psi_{\textrm{EF}}(r),
\end{equation}
where in contrast to the function $\psi(r)$,
\begin{equation}
\psi_{\textrm{EF}}(r) = e^{-\Omega(s)z}
\mbox{HeunC}\left( 2\Omega(s),4M_{\textrm{BH}} s,0,\delta,-\delta,-z \right)
\label{Eq:psiCloseBHEF}
\end{equation}
is now regular at the horizon, $z = 0$. The expressions for the metric coefficients,
\begin{subequations}
\begin{eqnarray}
 \alpha_{\textrm{EF}} &=& \gamma_{\textrm{EF}}^{-1} = \left(1+\frac{2M_{\textrm{BH}}}{r}\right)^{-1/2},\\
 \nu_{\textrm{EF}} &=& \frac{2M_{\textrm{BH}}}{r}\alpha_{\textrm{EF}},
\end{eqnarray}
\end{subequations}
as well as the scalar field energy,
\begin{equation}
E_{\textrm{EF}}(t_{\textrm{EF}}) = 4\pi\int\limits_{2m}^\infty 
\left( \rho - \frac{2M_{\textrm{BH}}}{r} j \right) r^2 dr,
\end{equation}
are now also manifestly regular, where
\begin{eqnarray}
 \rho - \frac{2M_{\textrm{BH}}}{r} j = \frac{1}{2} e^{2\sigma t_{\textrm{EF}}}\left\{
 \left( 1 - \frac{2M_{\textrm{BH}}}{r} \right)|\psi_{\textrm{EF}}'|^2 \right. &&
\nonumber\\
 + \left. \left[ \mu^2 + \left( 1 + \frac{2M_{\textrm{BH}}}{r} \right) |s|^2 \right] |\psi_{\textrm{EF}}|^2 \right\}. &&
\label{Eq:rhoEF}
\end{eqnarray}
Applying the balance law~(\ref{Eq:BalanceLaw}) to the solution in Eq.~(\ref{Eq:psiCloseBHEF}), one can simplify the expression for the energy to
\begin{equation}\label{eq:energy}
E_{\textrm{EF}}(t_{\textrm{EF}}) 
 = 4\pi\frac{(2M_{\textrm{BH}}|s|)^2}{|\sigma|}  e^{-2|\sigma| t_{\textrm{EF}}},
\end{equation}
which shows that $E_{\textrm{EF}}(t_{\textrm{EF}})$ decays exponentially as $t_{\textrm{EF}} \to +\infty$. This decay is due to the slow accretion of the scalar field mode into the black hole. Correspondingly, along the past direction $t_{\textrm{EF}}\to -\infty$, the energy $E_{\textrm{EF}}(t_{\textrm{EF}})$ diverges, which explains why the energy 
$E(t)$ measured on constant Schwarzschild time slices diverges.

Clearly, since we are neglecting the gravitational backreaction of these objects, the quasi-bound states described in this subsection only make physical sense if their amplitude is small enough (although at the mathematical level the solutions given in Eqs.~(\ref{Eq:psiCloseBH}) and~(\ref{Eq:psiCloseBHEF}) can be rescaled by an arbitrary factor since the differential equation~(\ref{eq:qbs}) is linear and homogeneous).
For large amplitudes, the gravitational effects associated with the scalar field need to be treated self-consistently by solving the coupled Einstein-Klein-Gordon equations. This is the main subject of the next section.

\section{Self-gravitating scalar wigs surrounding black holes}
\label{Sec:Results}

After reviewing the main properties concerning static boson starts and quasi-bound states surrounding Schwarzschild black holes, here we present the main new results of this work,
namely the extension of the quasi-bound states to the self-gravitating regime. After presenting our semi-analytic approximation 
method for constructing these modes, we provide numerical results and analyze the properties of these solutions. As expected, and as we also show explicitly here, these solutions reduce to the quasi-bound states in the limit in which the ratio between the masses of the scalar field configuration and the black hole is small.
In the opposite regime, that is, when the black hole mass is small compared to the total mass in the system, we show that these objects approach standard static boson star 
configurations. The results of this section will be corroborated with numerical evolutions in Sec.~\ref{Sec:Numerical}.

\subsection{The approximation method}
\label{Sec:Approximation}

Our approximation method is based on the same ansatz in Eq.~(\ref{Eq:Coherent}) that leads to the static boson stars and the quasi-bound states of Sec.s~\ref{Sec:boson stars} 
and~\ref{Sec:Quasi-bound states}, respectively. 
Like in the quasi-bound case but unlike in the boson star one, we now assume $s$ to have a negative real part $\sigma < 0$, that is, the scalar field decays exponentially in time. 
However, unlike the quasi-bound case but like in the boson star one, we now consider the full coupled Einstein-Klein-Gordon system in Eqs.~(\ref{eq:ESS}) and~(\ref{eq:dyn.scalar}).

In order to obtain the relevant equations in this section, we assume that the scalar field $\Phi$ and its canonical momentum $\Pi$ can be described, approximately, by a coherent state of the form
\begin{equation}
\left( \begin{array}{c} \Phi(t,r) \\ \Pi(t,r) \end{array}Ê\right) = e^{st}
\left( \begin{array}{c} \psi(r) \\ \pi(r) \end{array}Ê\right).
\label{Eq:CoherentBis}
\end{equation}
Plugging this ansatz into Eqs.~(\ref{Eq:mprime}), (\ref{Eq:D0nu}), and~(\ref{eq:dyn.scalar}), and adopting the $\nu = 0$ gauge yields the following system 
for the Misner-Sharp mass $m$, lapse $\alpha$, and radial wave $\psi$, functions:
\begin{widetext}
\begin{subequations}\label{Eq:bosonclouds}
\begin{eqnarray}
 m' = \frac{\kappa r^2}{4\gamma^2} e^{2\sigma t}
\left[ |\psi'|^2 + \gamma^2\left( \mu^2 + \frac{|s|^2}{\alpha^2} \right)|\psi|^2 \right],&&
\label{Eq:TIm}\\
 \frac{\alpha'}{\alpha} = \gamma^2\frac{m}{r^2} + \frac{\kappa r}{4} e^{2\sigma t}
\left[ |\psi'|^2 - \gamma^2\left( \mu^2 - \frac{|s|^2}{\alpha^2} \right)|\psi|^2 \right],&&
\label{Eq:TIalpha}\\
 \frac{1}{\alpha\gamma r^2}\left( r^2\frac{\alpha}{\gamma}\psi' \right)'
- \left[ \mu^2 + \frac{s^2}{\alpha^2} + \frac{\kappa r}{2\alpha^2} e^{2\sigma t} s\re(s\psi\psi'^*)
\right]\psi = 0,&&
\label{Eq:TIpsi}
\end{eqnarray}
\end{subequations}
\end{widetext}
with $\gamma = (1 - 2m/r)^{-1/2}$, and where we have used the fact that $K = -\kappa r\gamma j/2$. These are the equations describing our boson cloud configurations with a black 
hole in their center, analogous to those in Eqs.~(\ref{Eq:bosonstars}) for the case of an isolated static boson star (note that indeed these equations reduce to those in 
Eq.~(\ref{Eq:bosonstars}) when the frequency is real, $- is = \omega$, and $\psi$ is real-valued). Like in the boson star case, at spatial infinity we still impose 
asymptotic flatness conditions in order to obtain localized objects with finite mass.
However, now the regularity conditions at the origin $r=0$ described 
in Eqs.~(\ref{Eqs:BSOrigin}) need to be replaced with appropriate horizon boundary conditions which incorporate the presence of the black hole. 

This will be discussed next; however, before doing so, it is important to emphasize the following point: the remaining Einstein equation~(\ref{Eq:D0m}) yields
\begin{equation}
\dot{m} = \frac{\kappa r^2}{2\gamma^2} e^{2\sigma t}\re(s\psi\psi'^*),
\end{equation}
which shows that the metric cannot be strictly static, since $\sigma = \re(s) < 0$. This is also evident from Eqs.~(\ref{Eq:TIm}) and~(\ref{Eq:TIalpha}), due to the presence of 
the time-dependent factors $e^{2\sigma t}$ in their right-hand sides. Therefore, our method consists in finding solutions of the system in Eq.~(\ref{Eq:bosonclouds}) in which the 
factors $e^{2\sigma t}$ is set to one, for $r$-dependent functions $m$, $\alpha$, and $\psi$, and interpret them as approximate solutions of the Einstein-Klein-Gordon system for which the metric coefficients $m$ and $\alpha$ are time-independent while the scalar field has the time-dependent, slowly decaying form in Eq.~(\ref{Eq:Coherent}). 
Although this clearly does not yield an exact solution, it should provide a reasonably good approximation, at least for time scales $0\leq t \ll 1/\sigma$.\footnote{Time-dependent corrections to the metric can be considered by expanding the metric field in powers of $e^{2\sigma t}$, but this will be left to future investigation.} Alternatively, instead of thinking of the solutions of the system in 
Eq.~(\ref{Eq:bosonclouds}) as representing {\em approximate spacetime} solutions of the Einstein-Klein-Gordon system, we may use them as {\em exact initial data} for the 
full coupled field equations by defining the extrinsic curvature as discussed towards the end of Sec.~\ref{Sec:EoM}. In fact, this is what we will do in Sec.~\ref{Sec:Numerical} in order to explicitly check the validity of our approximate spacetime solutions.

After these remarks regarding the nature and validity of our approximation, we now focus our attention to the horizon boundary conditions. Close to the black hole, the boson 
cloud configurations are expected to have approximately the same form as a quasi-bound state, since in this region the gravitational potential is dominated by the black 
hole and the metric tensor is expected to be approximately the Schwarzschild one, at least locally. Therefore, we assume that sufficiently close to the inner boundary the scalar 
field is accurately described by a Klein-Gordon field propagating on a fixed Schwarzschild background of positive mass $m_0 = M_{\textrm{BH}}$, say, and thus the function $\psi(r)$ should 
have the form given in Eq.~(\ref{Eq:psiCloseBH}). However, as discussed in Sec.~\ref{Sec:Quasi-bound states}, this expression diverges as 
$r\to r_0 := 2M_{\textrm{BH}}$, and further the energy of the scalar field also diverges in this limit, yielding an infinite contribution to the mass function. 
Therefore, it is not possible to impose horizon boundary conditions at the inner boundary $r = r_0$, since both $\psi$ and $m$ diverge there.

For this reason, our strategy is to impose horizon boundary conditions at some radius $r_1 = r_0(1 + z_1)$ slightly larger than $r_0$, where the value of the function $\psi$ 
and its first derivative are determined according to the expression in Eq.~(\ref{Eq:psiCloseBH}). More precisely, the horizon boundary condition for the system
in Eq.~(\ref{Eq:bosonclouds}) are:
\begin{subequations}\label{Eq:bosoncloudHorizon}
\begin{eqnarray}
m(r_1) &=& M_{\textrm{BH}} + \Delta m(r_1),\\
\alpha(r_1) &=& \sqrt{1-\frac{2m(r_1)}{r_1}},\\
\psi(r_1) &=& A z_1^{2M_{\textrm{BH}} s} f(z_1),\\
\psi'(r_1) &=& \left[ \frac{s}{z_1} + \frac{1}{r_0}\frac{d\log f}{dz}(z_1) \right] \psi(r_1),
\end{eqnarray}
\end{subequations}
with a non-vanishing constant $A$ and where $z_1 = r_1/r_0 - 1$ and the function $f$ is given by
\begin{equation}
f(z) = e^{-\Omega(s)z}\mbox{HeunC}\left( 2\Omega(s),4M_{\textrm{BH}} s,0,\delta,-\delta,-z \right).\label{Eq:fder}
\end{equation}
These conditions replace those in Eqs.~(\ref{Eqs:BSOrigin}) when a black hole is present. As in the case of boson stars, the value of the lapse function $\alpha$ at $r = r_1$ is unimportant, because the system~(\ref{Eq:bosonclouds}) with $t=0$ is still invariant with respect to the rescaling $(\alpha,s)\mapsto \lambda(\alpha,s)$ by a positive constant $\lambda$. Furthermore, since this system is also invariant with respect to a global phase shift transformation, $\psi\mapsto e^{i\varphi}\psi$, it is sufficient to take the constant $A$ real and positive. For small enough values of $A$ one should (and one does indeed, as shown further below) reproduce the quasi-bound states, since in this case the self-gravity of the scalar field can be neglected. However, in the method described in this section there is no restriction on the magnitude of $A$, except that the larger its value, the closer to the black hole horizon we should fix the interior boundary conditions, such that the quantity $\Delta m(r_1)/M_{\textrm{BH}}$ remains small.

\begin{figure*}
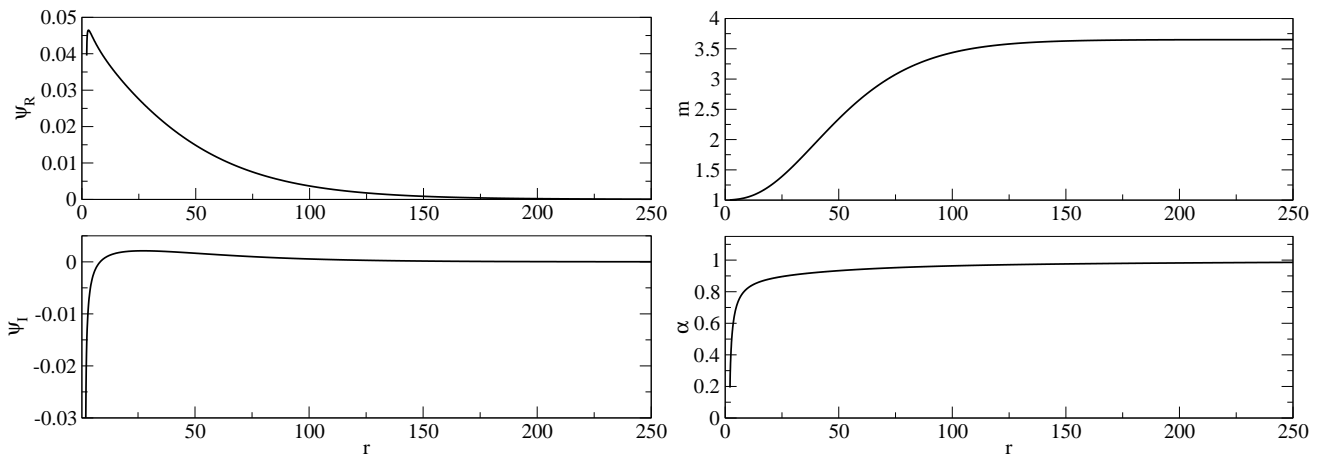

\includegraphics[angle=0,width=0.49\textwidth,clip]{psi3.eps}
\includegraphics[angle=0,width=0.47\textwidth,clip]{fmetricas.eps}
\caption{The real and imaginary parts of the wave function $\psi(r)$, the Misner-Sharp mass $m(r)$, and the lapse function $\alpha(r)$, for a black hole scalar wig with $A=0.0354$, $M_{\textrm{BH}}=1$, and $\mu=0.1$. Here the inner boundary has been
chosen at $r_1=2.1M_{\textrm{BH}}$.  See the fourth line in Tab.~\ref{Table1} for more details about this self-gravitating object.}
\label{Fig2}  
\end{figure*}

Now we address the problem of the computation of $\Delta m(r_1)$. Under our assumption wherein the self-gravity of the scalar field may be neglected close to the black hole 
horizon, we have seen in Sec.~\ref{Sec:Quasi-bound states} that while the energy $E$ 
diverges on constant Schwarzschild time slices, it is finite when computed on 
spacelike slices which penetrate the future horizon. Therefore, we can first compute the change of mass $\Delta m$ due to the scalar field in iEF coordinates $(t_{\textrm{EF}},r)$ and then change the result to the gauge $\nu=0$ we are using in this section. 

Given the value for $\Delta m$ at some point $(t_{\textrm{EF}},r) = (0,r_0)$, say, 
the value for $\Delta m$ at any other point can be obtained using the integral formula~(\ref{Eq:mIntegral}) in Appendix~\ref{App:ConservationLaws}. Noticing that in the test 
field limit $\psi_{\textrm{EF}}(r) = z^{-2M_{\textrm{BH}} s} \psi(r) = A f(z)$, we obtain
\begin{widetext}
\begin{equation}
 \Delta m(t_{\textrm{EF}},r_1) = \Delta m(0,r_0) 
 + 2\delta \int\limits_{r_0}^{r_1} \left( \rho - \frac{2M_{\textrm{BH}}}{r} j \right) r^2 dr
 + \delta\frac{|s|^2}{|\sigma|} r_0^2\left( 1 - e^{2\sigma t_{\textrm{EF}}} \right),
\label{Eq:Deltam1}
\end{equation}
where we have set $\delta := \kappa A^2/4$, and where $\rho - 2M_{\textrm{BH}} j/r$ is given by Eq.~(\ref{Eq:rhoEF}) in which $\psi_{\textrm{EF}}$ is replaced by 
$\psi_{\textrm{EF}}(r) = f(z)$. The integral expression on the right-hand side can be further simplified by applying the balance law~(\ref{Eq:BalanceLaw}) to the solution in Eq.~(\ref{Eq:psiCloseBHEF}), which yields the identity
\begin{eqnarray}
&& \int\limits_{r_0}^{r_1} \left\{
 \left(1 - \frac{r_0}{r} \right)|f'(r)|^2 + \left[ \mu^2 
 + \left(1 + \frac{r_0}{r} \right)|s|^2 \right] |f(r)|^2 \right\} r^2 dr
 = \nonumber \\
&& \frac{r_0^2}{|\sigma|} |s|^2 |f(r_0)|^2 - \frac{r^2}{|\sigma|}\left[
 \frac{r_0}{r} |s|^2 |f(r)|^2 + \left(1 - \frac{r_0}{r} \right)\re(s f f'^*) \right]_{r=r_1}.
\end{eqnarray}
Therefore, the integral in Eq.~(\ref{Eq:Deltam1}) can be eliminated completely and we end up with the simple expression
\begin{equation}
\Delta m(t_{\textrm{EF}},r_1) = \Delta m(0,r_0) + \frac{\delta}{|\sigma|}\left\{  r_0^2 |s|^2 
 - e^{2\sigma t_{\textrm{EF}}} r_1^2\left[ 
 \frac{r_0}{r} |s|^2 |f|^2 + \left(1 - \frac{r_0}{r} \right)\re(s f f'^*) \right]_{r=r_1} \right\}
\label{Eq:Deltam}
\end{equation}
\end{widetext}
for the change of the mass function at the point $(t_{\textrm{EF}},r_1 > r_0)$ due to the presence of the scalar field. In the following, we set the integration constant 
$\Delta m(0,r_0)$ to zero, which implies that at iEF time $t_{\textrm{EF}} = 0$ the apparent horizon has areal radius $r_0$: $2m(t_{\textrm{EF}} = 0,r_0) = r_0$. Notice that in this case, the correction $\Delta m(t_{\textrm{EF}},r_1)$ is small for $t_{\textrm{EF}}\geq 0$ as long as $\delta/|\sigma| \ll 1$ and $r_1$ lies close to $r_0$.

Finally, we transform this formula back to a constant Schwarzschild time slice by setting $t_{\textrm{EF}} = 2M_{\textrm{BH}}\log(z)$, such that
$$
\Delta m(r) := \left. \Delta m(t_{\textrm{EF}},r) \right|_{t_{\textrm{EF}} 
 = 2M_{\textrm{BH}}\log\left( \frac{r}{2M_{\textrm{BH}}} - 1 \right)}
$$
in the horizon conditions in Eq.~(\ref{Eq:bosoncloudHorizon}). This concludes the discussion of our horizon boundary conditions, and we are now ready to solve the system in 
Eq.~(\ref{Eq:bosonclouds}).

\subsection{Results from a numerical shooting algorithm}\label{sec.algorithm.bhsh}

The system of Eqs.~(\ref{Eq:bosonclouds}) with $e^{2\sigma t}=1$ and boundary conditions as expressed in Eqs.~(\ref{Eq:bosoncloudHorizon}) can be solved for any value of $s$. Nevertheless, since we are mainly interested in self-gravitating compact objects, we need to search for those values of $s$ for which the scalar field decreases 
asymptotically to zero at spatial infinity, such that the mass $m$ and the lapse $\alpha$ converge to a constant value at large radii. Again, this constitutes a nonlinear eigenvalue problem for the complex frequency~$-i s$.

In order to proceed, we follow an approach similar to that outlined in Sec.~\ref{Sec:boson stars} for the boson star case, except that now we start the integration from the inner boundary at $r = r_1$ where the horizon boundary conditions~(\ref{Eq:bosoncloudHorizon}) are imposed, and integrate towards an outer boundary at a large but finite value of the radial coordinate, $r = r_{\textrm{max}}$, using a second order shooting method. For configurations which are localized in space, it follows from Eqs.~(\ref{Eq:bosonclouds}) that the scalar field must decay exponentially at spatial infinity, more specifically, it should behave as $\psi(r) \sim \exp [ -\sqrt{\mu^2+s^2/\alpha^2}\,r]$ for $r\to\infty$.\footnote{More precisely, the asymptotic behavior of the scalar field is given by $\psi(r) \sim (r/2M)^{-p} e^{-q r}$, with $p = 1 + 2M q - M\mu^2/q$, $q =  \sqrt{\mu^2+s^2/\alpha_\infty^2}$, and $M = m_\infty$ the total mass. However, this only adds small terms of the order $1/r$ to the right-hand sides of Eqs.~(\ref{Eqs:BSAsymptotic3}), (\ref{Eq:asymptoticRe}) and~(\ref{Eq:asymptoticIm}), which can be neglected provided $r_{\textrm{max}}$ is chosen large enough.} This behavior is 
analogous to that of a static boson star; however, now the function $\psi(r)$ and the frequency $- i s$ are in general complex. Accordingly, the outer boundary conditions in Eqs.~(\ref{Eqs:BSAsymptotic2}) are generalized to
\begin{subequations}\label{Eqs:BSAsymptotic3}
\begin{eqnarray}
 \psi(r_{\textrm{max}}) &=& \psi_m,\\
 \psi'(r_{\textrm{max}}) &=& -\sqrt{\mu^2 + \frac{s^2}{\alpha^2}}\,\psi_m.
\end{eqnarray}
\end{subequations}
This translates into two different conditions for the real, 
\begin{subequations}\label{Eq:asymptoticRe}
\begin{eqnarray}
\psi_R(r_{\textrm{max}})&=&\psi_R^m,\\
\psi_R'(r_{\textrm{max}})&=&-\lambda_R \psi^m_R+\lambda_I \psi^m_I,
\end{eqnarray}
\end{subequations}
and imaginary parts 
\begin{subequations}\label{Eq:asymptoticIm}
\begin{eqnarray}
\psi_I(r_{\textrm{max}})&=&\psi_I^m,\\
\psi_I'(r_{\textrm{max}})&=&-\lambda_R \psi^m_I-\lambda_I \psi^m_R,
\end{eqnarray}
\end{subequations}
of the wave function, where $\lambda:=\sqrt{\mu^2+s^2/\alpha^2}$, and since $s$ is complex, $\lambda$ is a complex number as well, i.e. $\lambda=\lambda_R+i\lambda_I$.
For any given value of the field amplitude $A$, there is an infinite tower of possible states $s_n(A)$ satisfying the system~(\ref{Eq:bosonclouds}) together with the boundary 
conditions in Eqs.~(\ref{Eq:bosoncloudHorizon}) and~(\ref{Eqs:BSAsymptotic3}). However, in the following we will focus our attention on the ground state only, that is, the state 
for which the magnitude of $\omega$ is minimal. These self-gravitating objects are referred to as black hole scalar wigs in this paper.

The behavior of the different physical quantities in a typical black hole scalar wig is exhibited in Fig.~\ref{Fig2}, where the real and imaginary parts of the wave function, $\psi(r)$, as well as the Misner-Sharp mass, $m(r)$, and lapse function, $\alpha(r)$, are shown for 
a combination of the parameters such that $M_{\textrm{BH}}=1$, $\mu = 0.1$,  
and $A = 0.0354$. Here the inner boundary has been chosen at $r_1 = 2.1M_{\textrm{BH}}$, close enough to the apparent horizon, such that the final results are not substantially affected by the choice of $r_1$. 
The scaling symmetry of the Einstein-Klein-Gordon system that has been mentioned previously now implies
\begin{eqnarray}\label{eq.sym3}
 \mu\to a\mu,\quad s\to as,\quad r\to a^{-1}r, \nonumber\\
 M_{\textrm{BH}}\to a^{-1}M_{\textrm{BH}},\quad  m\to a^{-1}m,
\end{eqnarray}
with $\psi$, $\alpha$ and $\gamma$ unchanged. This makes it possible to map given solutions in the parameter space $(M_{\textrm{BH}},\mu,A)$ to new ones, keeping the product $M_{\textrm{BH}}\mu$ and field amplitude $A$ fixed. This feature will turn out to be 
essential in the discussion section, when making contact with real astrophysical objects.

Note that the value of $A$ affects the local solution determining the horizon boundary conditions in Eqs.~(\ref{Eq:bosoncloudHorizon}). 
In particular, increasing the field amplitude gives rise to solutions with higher ratios $M_{\textrm{T}}/M_{\textrm{BH}}$ between the total and black hole masses, where $M_{\textrm{T}}$ is defined as the value of the Misner-Sharp mass function at the final numerical integration point $r_{\textrm{max}}$, 
\begin{equation}
M_{\textrm{T}}:=m(r_{\textrm{max}}).
\end{equation}
In analogy with static boson stars this behavior is expected at least until a critical value of $A$ is reached, beyond which the ratio $M_{\textrm{T}}/M_{\textrm{BH}}$ starts to decrease and the system becomes unstable.
We leave a more detailed analysis regarding this conjecture for future work. Tab.~\ref{Table1} shows the relevant quantities for some self-gravitating configurations with 
different values of $A$, again with $M_{\textrm{BH}}=1$ and $\mu=0.1$.

\begin{figure}
\includegraphics[angle=0,width=0.48\textwidth,clip]{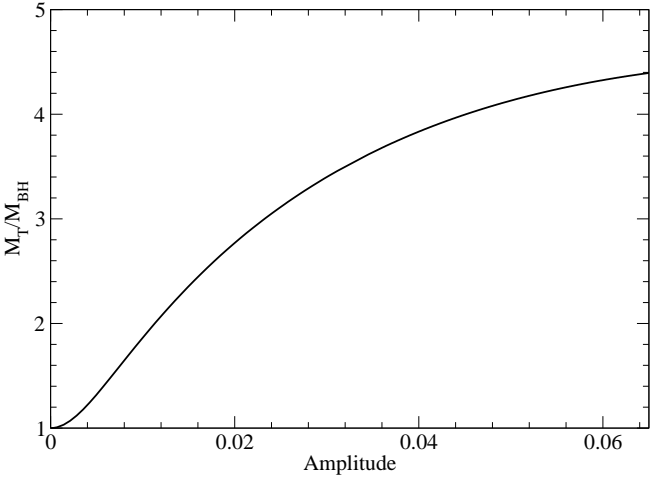}
\caption{The total to black hole mass ratio, $M_{\textrm{T}}/M_{\textrm{BH}}$, of a black hole scalar wig of $M_{\textrm{BH}}=1$ and $\mu=0.1$ as function of the field amplitude $A$. 
}\label{Fig3}
\end{figure}

\begin{table}
\begin{tabular}{c|c|c|c}
\hline
$A$ &$\sigma$&$\omega$& $M_{\textrm{T}}/M_{\textrm{BH}}$ \\
\hline
$A_1=7.71\times 10^{-7}$ & $-1.533\times 10^{-5}$ & $9.944\times 10^{-2}$ & 1.00001\\
$A_2=5.66\times 10^{-3}$ & $-2.041\times 10^{-5}$ & $9.911\times 10^{-2}$ & 1.39012\\
$A_3=1.14\times 10^{-2}$ & $-3.196\times 10^{-5}$ & $9.845\times 10^{-2}$& 2.00696\\
$A_4=3.54\times 10^{-2}$ & $-1.205\times 10^{-4}$ & $9.608\times 10^{-2}$ & 3.65115\\
$A_5=6.36\times 10^{-2}$ & $-3.282\times 10^{-4}$ & $9.256\times 10^{-2}$ & 4.37737\\
\hline
\end{tabular}
\caption{Decay rates $\sigma$, frequencies $\omega$, and total to black hole mass ratios $M_{\textrm{T}}/M_{\textrm{BH}}$, for some self-gravitating configurations of different field amplitudes $A$, but same values of $M_{\textrm{BH}}=1$ and $\mu=0.1$. Here the frequencies and decay rates have been normalized in such a way that the lapse function is fixed to one at spatial infinity. See Fig.~\ref{Fig3} for a corresponding plot of $M_{\textrm{T}}/M_{\textrm{BH}}$ vs. $A$.}
\label{Table1}
\end{table} 

\begin{figure}
\includegraphics[angle=0,width=0.48\textwidth,clip]{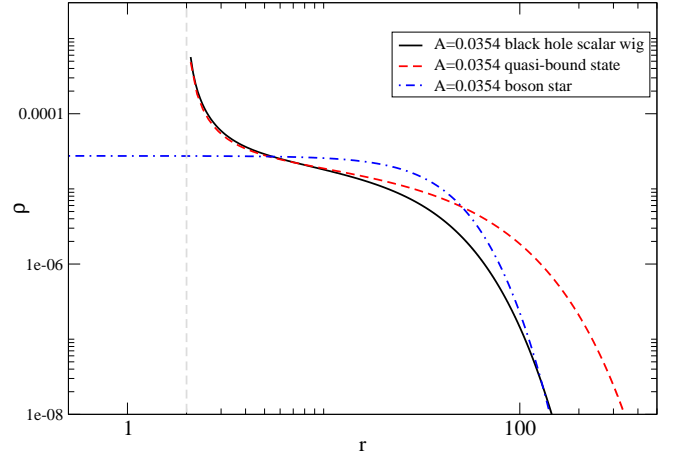}
\caption{The energy density $\rho$ as defined in Eq.~(\ref{eq.energydensity}) for different configurations of same field amplitude~$A$.
Note that the black hole scalar wig appears more compact than the quasi-bound state. At large radii the presence of the central black hole is not relevant
and the black hole scalar wig and the static boson star approach each other.
}\label{Fig5}
\end{figure}

As may be appreciated from Fig.~\ref{Fig3}, the total mass $M_{\textrm{T}}$ of these configurations with fixed parameters $M_{\textrm{BH}}=1$ and $\mu=0.1$ but varying field 
amplitude $A$ we have constructed can have their mass as large as several times the black hole mass, so the self-gravity of the corresponding objects is important. 
For these parameter choices, they have larger decay rates than their associated quasi-bound states with negligible self-gravity, as can be noticed for instance in Tab.~\ref{Table1}.
The presence of the black hole also compresses the scalar field configurations, making the self-gravitating objects more compact than the corresponding quasi-bound state. 
This can be seen in Fig.~\ref{Fig5}, where we compare the energy density of a quasi-bound state (i.e. no backreaction included), with the corresponding self-gravitating object of 
same amplitude. For the sake of illustration we also show the profile of a static boson star in this figure, which is regular at the origin.

\subsection{Boson stars and quasi bound states as a limit of the black hole scalar wigs}\label{subsec.limits}

\begin{figure}
\includegraphics[angle=0,width=0.48\textwidth,clip]{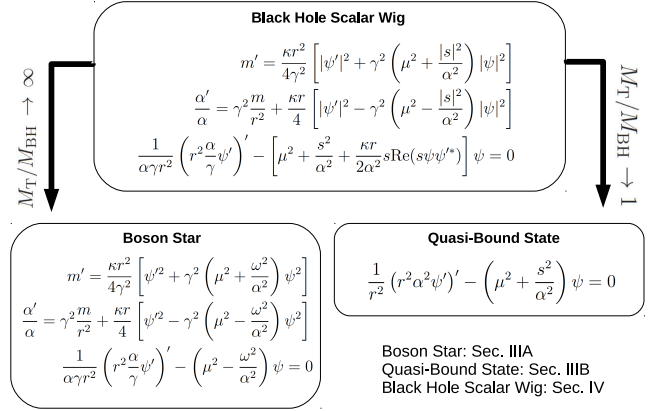}
\caption{The different limits of a black hole scalar wig. If the mass of the black hole tends to zero, static boson stars are recovered. 
As long as the field amplitude remains small, the quasi-bound states are a good approximation.}
\label{Fig:VariosLimites}
\end{figure}

As discussed previously, if the wave function $\psi(r)$ and the mode frequency are chosen to be real, then the nonlinear eigenvalue problem of the previous subsection reduces to the Einstein-Klein-Gordon system describing the static boson stars of Sec.~\ref{Sec:boson stars}. If on the contrary we turn off the self-gravitational interaction of the scalar 
field, $\kappa=0$, then this same system of equations describes the quasi-bound states of Sec.~\ref{Sec:Quasi-bound states}. In what follows, we show explicitly how the black 
hole scalar wigs smoothly transit between boson star configurations and Schwarzschild black 
holes dressed with long-lived scalar test field distributions in appropriate limits of the model parameters. (See Fig.~\ref{Fig:VariosLimites} for a clarifying picture.)

In order to do so, we analyze a family of black hole scalar wigs characterized by the same value of the parameters $\mu$ and $A$, but different black hole masses $M_{\textrm{BH}}$. 
In particular, we fix $\mu=0.1$ and $A=5.66\times 10^{-3}$, and solve the system in Eqs.~(\ref{Eq:bosonclouds}) as explained in Sec.~\ref{sec.algorithm.bhsh}, starting
from $M_{\textrm{BH}}=0.06$, and increasing the value of the black hole mass up to $M_{\textrm{BH}}=1$. Each configuration in this family has well-defined outputs for the wave 
function $\psi(r)$ and mode frequency $- i s$, and in what follows we discuss the behavior of these two quantities as function of the total to black hole mass ratio $M_{\textrm{T}}/M_{\textrm{BH}}$.

\begin{figure*}
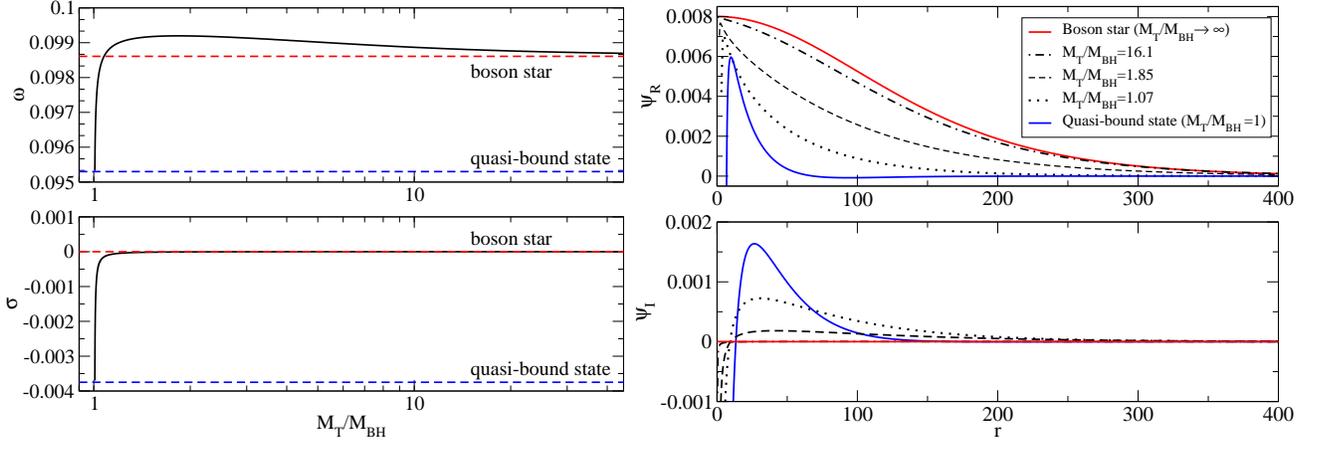

\includegraphics[angle=0,width=0.46\textwidth,clip]{transicion.eps}
\includegraphics[angle=0,width=0.49\textwidth,clip]{limitBS.eps}
\caption{Real and imaginary parts of $s=\sigma+i\omega$, as well as of the wave function $\psi(r)$, as a function of the ratio, $M_{\textrm{T}}/M_{\textrm{BH}}$, for a family
of black hole scalar wigs characterized by $\mu=0.1$, $A=5.66\times 10^{-3}$. 
Notice that a quasi-bound state and a static boson star are recovered in the limits $M_{\textrm{T}}/M_{\textrm{BH}}\to 1$ and 
$M_{\textrm{T}}/M_{\textrm{BH}}\to \infty$, respectively.
}\label{Fig.freq.limits}
\end{figure*}

In Fig.~\ref{Fig.freq.limits}, left column, we plot the real and imaginary parts of $s=\sigma+i\omega$ as a function of $M_{\textrm{T}}/M_{\textrm{BH}}$, for configurations in the above mentioned family. As expected, when the black hole dominates the self-gravitational configuration, i.e.
$M_{\textrm{T}}/M_{\textrm{BH}}\to 1$, the black hole scalar wig frequencies approach those associated with a quasi-bound state, in this case the one characterized by 
$M_{\textrm{BH}}=1$ and $\mu=0.1$. In the opposite limit, when the mass of the black hole is negligible compared to the total mass of the self-gravitating object, $M_{\textrm{T}}/M_{\textrm{BH}}\to \infty$, we recover the frequency of the static boson star with $\mu=0.1$ and $A=5.66\times 10^{-3}$. 
Note that this is a nontrivial result, because even if the black hole is so small that it barely contributes to the overall gravitational field in the exterior region, 
we cannot think of it simply as a small perturbation of the metric, due to the large compactness ratio $2m/r$ in the vicinity of the apparent horizon.

Similarly, in the right column of Fig.~\ref{Fig.freq.limits} we show the smooth transition along the aforementioned family of scalar wigs from the boson star configurations to the Schwarzschild 
black hole dressed with the quasi-bound state, now at the level of the wave function.

\section{Full nonlinear dynamical evolution and comparison}
\label{Sec:Numerical}

In order to validate our approximation scheme and to explore its limitations, 
in this section we perform numerical evolutions of the spherically symmetric 
Einstein-Klein-Gordon system. More specifically, we use the self-gravitating 
approximate configurations constructed in the previous section to provide 
initial data for the metric and the scalar field, which are then evolved 
numerically and compared to the approximate solutions. We show that the latter remain 
accurate for large amounts of time, supporting the validity of our approximation 
method.

In order to carry out the numerical evolutions,
we first rewrite the initial data in terms of horizon-penetrating coordinates.
More specifically, we change the time coordinate~$t$ associated with the 
gauge $\nu=0$ to an iEF one, defined as in Eq.~(\ref{Eq:tEFDef}), keeping the 
radial coordinate $r$ 
unchanged and $M_{\textrm{BH}}$ denoting the apparent horizon mass.
In these new coordinates the solutions to our approximation method in
Eq.~(\ref{Eq:Coherent}) take the form
\begin{eqnarray}\label{solutionnewcoordinates}
\Phi(t,r) &=& 
e^{st_{\textrm{EF}}}(r/2M_{\textrm{BH}}-1)^{-2M_{\textrm{BH}}s}\psi(r),
\end{eqnarray}
with the function $\psi(r)$ constructed as described in 
Sec.~\ref{sec.algorithm.bhsh} on the interval $[r_1, r_{\textrm{max}}]$. 
In order to extend this function on the computational domain 
$[0,r_{\textrm{max}}]$ we linearly interpolate it between $\psi(0) = 0$ and the 
value it assumes at $r = r_1$. Although this interpolation is somehow crude, it 
occurs entirely within the horizon and thus does not affect the 
solution in the exterior region.

For the evolution of the initial data we adopt the 
Baumgarte-Shapiro-Shibata-Nakamura formalism of 
Einstein's equations in spherical 
symmetry~\cite{Brown:2009dd,Alcubierre:2010is}, 
in which the three-metric has the form
\begin{equation}\label{eq:metric_3p1}
dl^2 = e^{4\phi(t,r)}[a(t,r)dr^2 + b(t,r)r^2d\Omega^2 ].
\end{equation}
Here $e^\phi$ is a conformal factor and $a$ and $b$ are metric coefficients. 
It has been shown that this formulation is 
particularly suitable for evolving spacetimes with black 
holes~\cite{Alcubierre:2010is, Sanchis-Gual:2014ewa}. The gauge conditions we 
use are the \emph{1+log} condition for the lapse and a \emph{Gamma-driver} 
condition for the shift vector in order to avoid the slice stretching of 
coordinates.
Although the dynamical gauge conditions adopted here differ from the ones 
described in previous sections, they have become standard in the numerical 
evolution of black hole spacetimes and thus we adopt them here as well, 
providing an extra test to validate the findings in Sec.~\ref{Sec:Results}.

\begin{figure}
\includegraphics[angle=0,width=0.48\textwidth,clip]{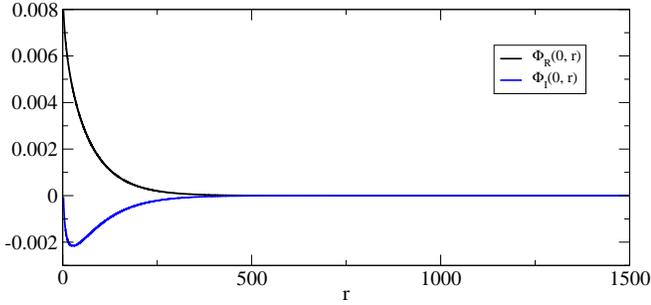}
\caption{Real and imaginary parts of the scalar field $\Phi$ as a function of 
$r$ at fixed iEF time $t_{\textrm{EF}} = 0$.}
\label{fig:phi_EF}
\end{figure}

The initial scalar field distribution described by 
Eq.~(\ref{solutionnewcoordinates}) at $t_{\textrm{EF}}=0$ is shown in 
Fig.~\ref{fig:phi_EF}. In order 
to obtain the initial metric and curvature coefficients, we solve the 
Hamiltonian 
and momentum constraints in spherical symmetry assuming, without loss of 
generality, 
that the initial slice is conformally flat, that is $a(0,r)=b(0,r)=1$.  
Under these assumptions the Hamiltonian and 
momentum constraints reduce to ordinary differential equations for the 
conformal factor, $e^{\phi}$, and the radial component of the 
extrinsic curvature. Furthermore, we write the conformal factor in a 
puncture-like form
\begin{equation}
e^\phi = 1 + \frac{M_{\rm BH}}{2r}+ u(r),
\end{equation}
assuming that $u(r)$ vanishes at spatial infinity. 
We solve the coupled equations using a fourth order Runge-Kutta 
integrator. Initially the shift is set to zero and we used a precollapsed lapse 
of the form
$\alpha = e^{-\phi}$.

We numerically evolve the Einstein-Klein-Gordon system for the models $A_1$ and 
$A_2$ of Tab.~\ref{Table1}. The first case constitutes a small deviation 
from the test field regime, since the scalar field only contributes little to 
the total mass of the system,
whereas in the second case the mass of the scalar configuration is much larger 
and comparable to that of the black hole. In both cases we found that the 
solution in Eq.~(\ref{solutionnewcoordinates}) accurately describes 
the evolution over large (coordinate) time spans; at least for time 
periods for which we were able to maintain the code stable 
($t\sim 1800$) before noise coming from the boundaries interferes with the 
field.

\begin{figure}
\includegraphics[angle=0,width=0.48\textwidth,clip]{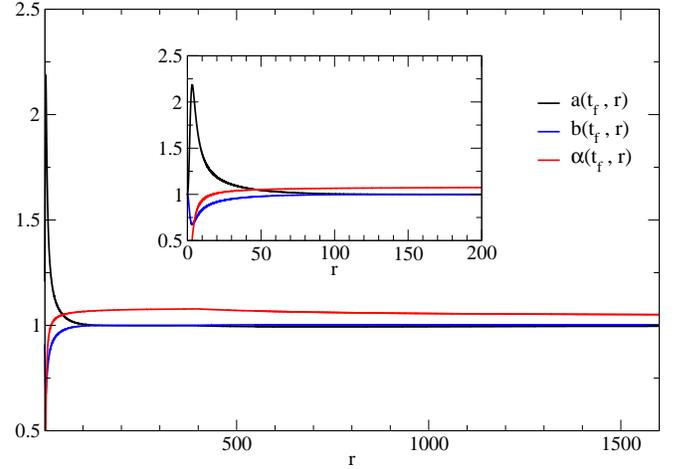}
\caption{Metric coefficients for $t_{\rm f}\sim 1800$ as a function of $r$.
The inset is a zoom onto the region in which most of the scalar field density is 
concentrated.
}
\label{fig:metric_coeff}
\end{figure}

\begin{figure}
\includegraphics[angle=0,width=0.48\textwidth,clip]{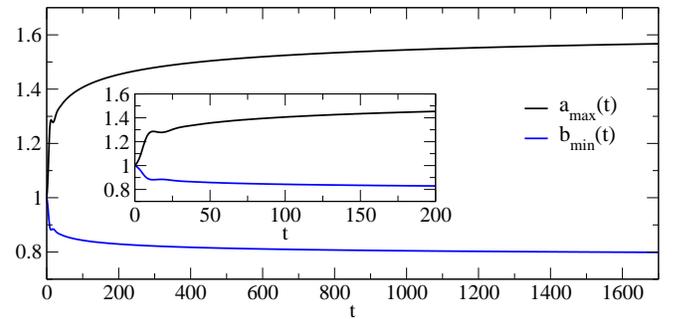}
\caption{Maximum (minimum) value of the metric coefficient $a$ ($b$). The inset 
shows the behavior at earlier times. At $t=0$ the spacetime is conformally flat 
and then it evolves towards a quasi-stationary state in the \emph{1 + log} 
slice.}
\label{fig:fix_metric}
\end{figure}

In Fig.~\ref{fig:metric_coeff} we plot the metric coefficients 
$a(t_{\rm f},r)$, $b(t_{\rm f},r)$ and $\alpha(t_{\rm f},r)$, where $t_{\rm 
f}\sim 1800$ is the final time of evolution. In Fig.~\ref{fig:fix_metric}
we show the evolution of the maximum (minimum) value of $a$ ($b$). 
The behavior of these two fields reflects the evolution of the metric. 
Initially, the metric is conformally flat and then it changes in time until it 
settles down to a new state which remains almost unchanged.

A Fourier transform of the maximum value of the real part of the scalar field 
shows that it oscillates according to Eq.~(\ref{solutionnewcoordinates}). 
The time of evolution is just large
enough to find the values of the frequencies $\omega_1=0.09943$ and 
$\omega_2=0.09914$ for the two cases
$A_1$ and $A_2$, respectively, that are in a good agreement with the values 
given in Tab.~\ref{Table1}. 

\begin{figure}
\includegraphics[angle=0,width=0.48\textwidth,clip]{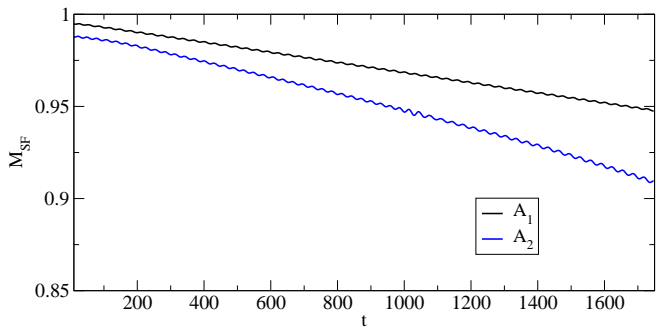}
\caption{Evolution of the scalar field mass for models $A_1$ and $A_2$.}
\label{fig:mass_SF}
\end{figure}

In Fig.~\ref{fig:mass_SF} we also monitor the mass of the scalar configuration 
throughout the evolution by integrating over the density given in 
Eq.~(\ref{eq:den_flux}) 
with $n^\mu = \alpha^{-1}(\delta^\mu{}_t - \beta\delta^\mu{}_r)$ 
the four-velocity of the Eulerian observers in 
the \emph{1+log} slice. We found the time decay rate via a 
numerical fit for this mass of the form $e^{2\sigma_n t}$. Our results indicate that  after 
an initial stage of complex dynamics, the mass presents an exponential decay, 
indicating that the field is falling into the black hole. For model $A_1$ we find a time rate decay $\sigma_n$ for the field of the
order $-1.5\times 10^{-5}$, whereas for model $A_2$ we obtain 
$-2\times 10^{-5}$, which are roughly the values of the predicted rates 
given in Tab.~\ref{Table1}. These results are consistent with the expression 
for the time evolution of the energy given by Eq.~(\ref{eq:energy}) in iEF 
coordinates. This result is interesting because it shows that once we have 
constructed 
the black hole scalar wigs on a regular foliation (the $t_{\textrm{EF}}=0$ 
slice) it remains finite in the other regular foliation determined by the 
\emph{1+log} slicing condition.

The main constraint in getting longer evolutions is the fact that the 
time step is 
limited by the Courant condition which requires it to be sufficiently smaller 
than the spatial grid size, which in turn needs to be small in order to keep 
the 
numerical solution stable. Thus, in our current setup, it is not possible to 
obtain evolutions for very large times with a spatial domain that is large 
enough such that the solution is not contaminated by boundary effects.

\section{Discussion and conclusions}
\label{Sec:Discussion}

In this paper we have extended the quasi-bound states to the regime where the gravitational backreaction of the scalar field is relevant. In order to do so, we introduced a new approximation method for semi-analytically solving the Einstein-Klein-Gordon system in presence of black holes. This method consists in proposing a coherent state ansatz for 
the scalar field which, together with appropriate (inner) horizon boundary conditions and asymptotic flatness at spatial infinity, leads to a nonlinear eigenvalue problem which is solved 
numerically. The viability of our approximation has been confirmed using numerical evolutions of the full nonlinear equations. 
Although in this work we have focused on nonrotating systems, it should in principle be possible to generalize it to more sophisticated scenarios including rotation.

Our results corroborate that, as expected from previous numerical dynamical evolutions~\cite{Okawa:2014nda,Sanchis-Gual:2014ewa}, the quasi-bound states possess analogues in the self-gravitating case, yielding self-consistent, asymptotically flat solutions of the 
Einstein-Klein-Gordon equations. We called these self-gravitating objects 
black hole scalar wigs, in analogy with the Schwarzschild scalar wigs described in our previous work~\cite{Barranco:2012qs, Barranco:2013rua}.

These objects are characterized by three independent parameters: the scalar field mass $\mu$ (like is the case for any solution of the Klein-Gordon equation), the black hole 
mass $M_{\textrm{BH}}$ (in analogy to the quasi-bound states), and a field amplitude $A$ (which plays a similar role to the central value of the scalar field in the static boson 
star case).\footnote{A further parameter in our setup is the position $r_1$ at which the inner boundary conditions are specified; however, as long as it is chosen close enough to the apparent horizon it does not affect the physical results.}
However, thanks to the scaling symmetry described in Eq.~(\ref{eq.sym3}), it is possible to eliminate one of the mass scales and characterize black hole scalar wigs solely in terms of two parameters.
It is convenient to express these two parameters in terms of the following dimensionless quantities: the product $M_{\textrm{BH}}\mu$ between the black hole and the scalar field 
masses (in Planck units), and the ratio $M_{\textrm{T}}/M_{\textrm{BH}}$ between the total and black hole masses.

As expected, the black hole scalar wigs change considerably as $M_{\textrm{T}}/M_{\textrm{BH}}$ increases, undergoing a transition from a quasi-bound state in the test field 
limit when $M_{\textrm{T}}/M_{\textrm{BH}} = 1$, to cases in which the mass of the scalar wig is several times larger than the mass of the black hole and represents a large 
boson star with a tiny black hole in its center (the black hole being tiny when compared
to the size of the whole self-gravitating object). In the context of the ultralight axion dark matter models these latter situations may represent the large core of the halos in presence of central supermassive black holes.

\begin{figure}
\includegraphics[angle=0,width=0.48\textwidth,clip]{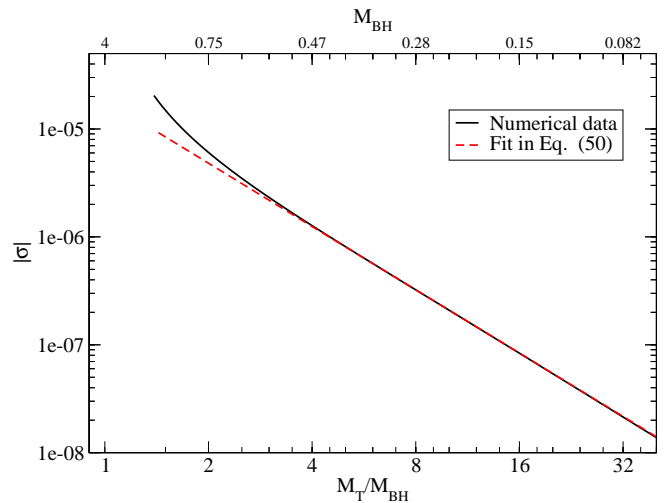}
\caption{Same plot as in Fig.~\ref{Fig.freq.limits} for the magnitude of the decay rate $|\sigma|$ as a function of the total to black hole mass ratio, now in a logarithmic scale. This plot makes the power-law behavior for $M_{\textrm{T}}/M_{\textrm{BH}}\gtrsim 3$ of this function evident.}
\label{fig.fit}
\end{figure}

In order to make contact with real astrophysical systems, we conclude with some rough estimates regarding the lifetime of the scalar field configurations constructed in this article. 
In order to do so, we first note that the half-lifetime of the black hole scalar wigs can be estimated from the results given in Sec.~\ref{Sec:Results},  
together with the expression in Eq.~(\ref{eq.timephys}).
Next, as a specific example we consider the family of black hole scalar wigs characterized by $\mu = 0.1$ and $A = 5.66\times 10^{-3}$. Fitting the magnitude of the decay rate 
as a function of the total to black hole mass ratio to a power law, see Fig.~\ref{fig.fit}, one obtains the empirical formula
\begin{subequations}\label{eq.fit}
\begin{equation}
 |\sigma| \approx a_1\left(M_{\textrm{T}}/M_{\textrm{BH}}\right)^{a_2},
\end{equation}
where
\begin{equation}
a_1= 1.8582\times 10^{-5},\quad
a_2= -1.9489.
\end{equation}
\end{subequations}
Introducing Eq.~(\ref{eq.fit}) into the expression~(\ref{eq.timephys}) yields
\begin{equation}
t_{1/2}[\textrm{years}]\sim \frac{1.1\times 10^3}{\mu[10^{-22}\,\textrm{eV}]}
\left(\frac{M_{\textrm{T}}}{M_{\textrm{BH}}}\right)^2.
\label{eq:t1}
\end{equation}
The black hole mass $M_{\textrm{BH}}$ in Planck units belonging to a particular configuration in the above family  can be read off, or extrapolated, from the upper horizontal 
axis in Fig.~\ref{fig.fit} and converted to physical units using the conversion formula 
$M_{\textrm{BH}}[10^8\,M_{\odot}]=1.3\times10^{3} M_{\textrm{BH}}/\mu[10^{-22}\,\textrm{eV}]$. As an example, considering a case where the scalar field configuration is hundred times higher than the black hole mass, Eq.~(\ref{eq:t1}) yields a time of the order of ten million years, corresponding to a black hole of mass $10^{9}M_{\odot}$.
Comparing this with the order $10^{13}$ years obtained from Eq.~(\ref{eq.t1/2test}) for a test field configuration with the same value for the product of 
$M_{\textrm{BH}}\mu\sim 10^3$, we notice that the self-gravitating configurations may have much smaller lifetimes,
although we stress that this conclusion is based on a very specific family of black hole scalar wigs.
We leave a more systematic analysis of these states and their applicability to real astrophysical systems for future work.

In practice, the self-gravitating scalar wigs we have discussed in this article may be detected through their dynamical influence on the stars, gas clouds, light rays or other compact objects surrounding the supermassive black hole at the center of galaxies. 
Indeed, the motion of these bodies is affected by all form of mass contained inside their orbital radius, that is, by the mass of any (hypothetical) matter distribution in addition to the black hole mass.
As we have shown, for the case of a scalar field, such distribution can be several times larger than the actual mass of the black hole. Near-future observations such as the Event 
Horizon Telescope project~\cite{Ricarte:2014nca} should yield accurate estimates for the size of the black hole shadow in Sagittarius A$^*$ and at the center of other galaxies, 
thus providing an estimate for the actual black hole mass. If there was a difference between this mass and the total mass measured through independent, dynamical observations, 
this would be a clear indication for the presence of large matter distributions surrounding the black hole, such as the ones discussed in the present work. 
A further possible signature for the presence of a scalar wig could be detected in the gravitational wave signal emitted by the accreted matter in the form of a monopole black 
hole ringdown~\cite{Okawa:2014nda}. However, we leave these and other studies for future investigation.


\acknowledgments
This work was supported in part by CONACYT grants No. 82787, 101353, 259228,
167335, and 182445, by the CONACYT Network Project 
280908 ``Agujeros Negros y Ondas Gravitatorias", by DGAPA-UNAM through grants 
IN103514 and IA103616, by SEP-23-005 through grant 181345, 
and by a CIC grant to Universidad Michoacana de San Nicol\'as de Hidalgo. A.B. was 
partially supported by PROMEP.

\appendix

\section{Conservation laws for the spherically symmetric Einstein-Klein-Gordon system}
\label{App:ConservationLaws}

The purpose of this appendix is to make a few general remarks about conservation laws for spherically symmetric systems, including the Einstein-Klein-Gordon one discussed in Sec.~\ref{Sec:EoM}. Although the conservation laws we discuss here are known, they are directly relevant for the results presented in this paper, and thus we summarize them for the convenience of the reader.

It is well-known that --unlike in the special relativistic case-- in general relativity the existence alone of a divergence-free stress-energy tensor $T_{\mu\nu}$ is not sufficient to construct conserved quantities from integrals of $T_{\mu\nu}$ over spacelike hypersurfaces $\Sigma$. The reason for this is that in a curved spacetime, it is in general not possible to integrate a vector field over $\Sigma$; hence, besides the future-directed unit normal vector $n^\mu$ to $\Sigma$, a further vector field $X^\mu$ is needed to contract with the stress-energy tensor $T_{\mu\nu}$. Given such a vector field $X^\mu$, one introduces the associated ``current density"
\begin{equation}
J_X^\mu := -T^\mu{}_{\nu} X^\nu,
\label{Eq:JXDef}
\end{equation}
which satisfies the divergence identity
\begin{equation}
\nabla_\mu J_X^\mu = S,\qquad
S := -T^{\mu\nu}\nabla_{(\mu} X_{\nu)},
\label{Eq:DivJX}
\end{equation}
where due to the symmetry of $T_{\mu\nu}$ only the symmetric part $\nabla_{(\mu} X_{\nu)} = (\nabla_\mu X_\nu + \nabla_{\mu} X_\nu)/2$ of the gradient of $X_\nu$ enters the ``source term" $S$. In general, $S$ does not vanish. However, when $S = 0$, then $J_X^\mu$ is divergence-free and in this particular case integration of Eq.~(\ref{Eq:DivJX}) over a spacetime region enclosed between two Cauchy surfaces and the application of Gauss' theorem leads to the ``conserved charge"
\begin{equation}
Q_X[\Sigma] = -\int\limits_\Sigma J_X^\mu n_\mu \sqrt{g_\Sigma} d^3x,
\label{Eq:conservedcharge}
\end{equation}
where $\Sigma$ denotes one of the Cauchy surfaces and $\sqrt{g_\Sigma} d^3x$ the induced volume element on it. (Here, we have assumed that the matter fields fall off sufficiently fast in the asymptotic region for the boundary terms to vanish. If $T_{\mu\nu}$ does not fall off sufficiently fast or if the spacetime region over which Eq.~(\ref{Eq:DivJX}) is integrated is enclosed between two spacelike hypersurfaces $\Sigma_{1,2}$ which are not Cauchy surfaces, then the difference between $Q_X[\Sigma_2]$ and $Q_X[\Sigma_1]$ is given by the boundary flux integrals, see further below for an example.) In particular, if $X^\mu$ is a future-directed timelike vector field satisfying $S = 0$ and if $T_{\mu\nu}$ satisfies the {\it dominant energy condition},\footnote{
The dominant energy condition states that for any future-directed 
timelike vector field $u^\mu$ the associated current density $J_u^\mu = -T^\mu{}_{\nu} u^\nu$ must be future-directed timelike or null~\cite{Wald84}.} then the conserved charge $Q_X$ is guaranteed to be nonnegative, and in this case it may be associated with ``mass" or ``energy".

A particular scenario for which $S = 0$ occurs when $X^\mu$ is a future-directed timelike Killing vector field, in which case $\nabla_{(\mu} X_{\nu)} = 0$ and the quantity $Q_X$  in Eq.~(\ref{Eq:conservedcharge}) is just the conserved energy $E$ associated with the time-invariance symmetry of the spacetime generated by $X^\mu$. However, dynamical spacetimes such as the ones analyzed in this work do not admit any timelike Killing vector fields and in this case one does not expect {\it a priori} the existence of a timelike vector field $X^\mu$ satisfying $\nabla_\mu J_X^\mu = 0$.

Interestingly, spherically symmetric spacetimes provide an exception to this expectation. In this case, even though the spacetime might be dynamical, there exists a natural choice for a vector field $X^\mu$ which does lead to a conserved current density $\nabla_\mu J_X^\mu = 0$. This vector field is known as the Kodama vector field~\cite{Kodama:1979vn}, see also~\cite{Racz:2005pm}. In order to describe it, we assume the spacetime to be oriented and denote by $\eta_{\alpha\beta\gamma\delta}$ the natural volume form induced by the spacetime metric $g_{\mu\nu}$. Denote by $\{ e_2^\mu, e_3^\nu \}$ a local oriented orthonormal frame tangent to the invariant two-spheres, and introduce the anti-symmetric tensor field
$$
\tilde{\eta}_{\alpha\beta} := \eta_{\alpha\beta\gamma\delta} e_2^\gamma e_3^\delta.
$$
This $\tilde{\eta}_{\alpha\beta}$ can be identified with the volume form on the two-dimensional manifold $\tilde{M}$ orthogonal to the invariant two-spheres. Then, the Kodama vector field is defined as
\begin{equation}
K^\mu := \tilde{\eta}^{\mu\nu}\nabla_\nu r.
\label{Eq:Kodama}
\end{equation}
By construction, the integral curves of $K^\mu$ leave $r$ invariant and thus $K^\mu$ is timelike in the region where the gradient of $r$ is spacelike, that is, in the region outside trapped spheres. As a consequence of Einstein's field equations it is not difficult to verify that $K^\mu$ satisfies
\begin{equation}
\nabla_\mu K_\nu + \nabla_\nu K_\mu
 = \kappa r\tilde{\eta}_{(\mu}{}^\alpha T_{\nu)\alpha},
\label{Eq:Killing}
\end{equation}
such that the Kodama vector field $K^\mu$ is a Killing field whenever $T_{\mu\nu} = 0$, or more generally, whenever the projection of $T_{\mu\nu}$ on $\tilde{M}$ is proportional to the induced metric on $\tilde{M}$. In terms of the ADM parametrization of the metric tensor discussed in Sec.~\ref{Sec:EoM}, we find
\begin{equation}
K^\mu = \frac{1}{\gamma}n^\mu + \nu w^\mu
 = \frac{1}{\alpha\gamma}\delta^\mu{}_t,
\label{Eq:JADM}
\end{equation}
which yields the timelike Killing vector $\partial/\partial t$ in the Schwarzschild case. Remarkably, even when $K^\mu$ is not a Killing vector field, it still yields a conserved current density when contracted with $T_{\mu\nu}$. Indeed, using Eq.~(\ref{Eq:Killing}) one finds
\begin{equation}
\nabla_\mu J_K^\mu = -T^{\mu\nu}\nabla_{(\mu} K_{\nu)}
 = -\frac{\kappa r}{2}\tilde{\eta}^{\mu\nu} T^\alpha{}_\mu T_{\alpha\nu} = 0,
\label{Eq:JK}
\end{equation}
and thus one obtains a conserved charge $Q_K$ of the form given in Eq.~(\ref{Eq:conservedcharge}). For asymptotically flat, spherically symmetric spacetimes, this charge coincides with the total (ADM) mass $M_{\textrm{ADM}}$ of the spacetime, as we show next.

More generally, the current density $J_K^\mu$ constructed from the Kodama vector field defined in Eq.~(\ref{Eq:Kodama}) is directly related to the Misner-Sharp mass function defined in Eq.~(\ref{Eq:MSMass}). In differential form, this relation is given by
\begin{equation}
\kappa r^2 J_K^\mu = -\tilde{\eta}^{\mu\nu}\nabla_\nu(2m),
\label{Eq:KodamaMS}
\end{equation}
which can be deduced from Eqs.~(\ref{Eq:JADM}) and the field equations~(\ref{Eq:D0m}) and~(\ref{Eq:mprime}) for the mass function $m$. (Note that the integrability condition $\nabla_{[\mu}\nabla_{\nu]}(2m) = 0$ for obtaining the mass function through an integral of  Eq.~(\ref{Eq:KodamaMS}) is equivalent to the conservation law $\nabla_\mu J_K^\mu = 0$.) Integrating Eq.~(\ref{Eq:JK}) over a rectangular region of spacetime of the form $[t_1,t_2]\times [r_1,r_2]\times S^2$ which is enclosed between the two partial Cauchy surfaces $\{ t_1 \}\times [r_1,r_2]\times S^2$ and $\{ t_2 \}\times [r_1,r_2]\times S^2$ and using Gauss' theorem, one finds the balance law
\begin{eqnarray}
\left. \int\limits_{r_1}^{r_2} r^2 T_{\mu\nu} K^\mu n^\nu \gamma dr \right|_{t_1}^{t_2}
= \left. \int\limits_{t_1}^{t_2} r^2 T_{\mu\nu} K^\mu (\gamma\nu n^\nu + w^\nu)\alpha dt
\right|_{r_1}^{r_2}.
\label{Eq:BalanceLaw}
\end{eqnarray}
This law could have equally well be obtained by integrating the Eqs.~(\ref{Eq:D0m}) and~(\ref{Eq:mprime}) for the mass function $m$, which, in terms of the Kodama vector field $K^\mu$ can be rewritten as
\begin{eqnarray}
\frac{\partial m}{\partial t} &=& 
\frac{\kappa r^2}{2} \alpha T_{\mu\nu} K^\mu (\gamma\nu n^\nu + w^\nu),
\label{Eq:D0mBis}\\
\frac{\partial m}{\partial r} &=& \frac{\kappa r^2}{2}\gamma T_{\mu\nu} K^\mu n^\nu.
\label{Eq:mprimeBis}
\end{eqnarray}
Combining Eqs.~(\ref{Eq:BalanceLaw}), (\ref{Eq:D0mBis}) and~(\ref{Eq:mprimeBis}) we also find the identity
\begin{eqnarray}
m(t_2,r_2) = m(t_1,r_1)
 &+& \frac{\kappa}{2}\left. \int\limits_{t_1}^{t_2} r^2 T_{\mu\nu} K^\mu (\gamma\nu n^\nu + w^\nu) \alpha dt \right|_{r=r_1}
\nonumber\\
 &+& \frac{\kappa}{2}\left. \int\limits_{r_1}^{r_2} r^2 T_{\mu\nu} K^\mu n^\nu \gamma dr \right|_{t=t_2},
\label{Eq:mIntegral}
\end{eqnarray}
which can be used to determine the mass function at an arbitrary point $(t_2,r_2)$ of the spacetime manifold from its value at a fixed reference point $(t_1,r_1)$ and an integral over the stress-energy tensor. In particular, when $t_2 = t_1 = t$ the identity~(\ref{Eq:mIntegral}) reduces to
\begin{equation}
m(t,r_2) = m(t,r_1) 
 + \frac{\kappa}{2}\left. \int\limits_{r_1}^{r_2} r^2 T_{\mu\nu} K^\mu n^\nu \gamma dr \right|_t,
\end{equation}
and as long as the dominant energy condition is fulfilled and $K^\mu$ is timelike it follows that $m(r_2,t)\geq m(r_1,t)$ for $r_2\geq r_1$, that is, the mass contained inside a spherical shell is positive. For an asymptotically flat spacetime the Misner-Sharp mass function $m(t,r)$ converges to the total mass $M_{\textrm{ADM}}$ when $r\to \infty$ along a hypersurface $t = \textrm{const.}$ which reaches spatial infinity. The identity~(\ref{Eq:mIntegral}) is key to our discussion in Sec.~\ref{Sec:Results}.

Finally, we note that for the particular case where the matter field is described by a canonical complex scalar field $\Phi$ with a $U(1)$-symmetric potential, there is an additional conserved current density which is associated with the internal $U(1)$ symmetry of the field and is given by
\begin{equation}
J_\mu = -\frac{i}{2}[\Phi(\nabla_\mu\Phi^*)-(\nabla_\mu\Phi)\Phi^*].
\end{equation}
The corresponding conserved charge $Q$ defined by Eq.~(\ref{Eq:conservedcharge}) is usually identified with the difference between the number of particles and antiparticles in the configuration. For the particular case of the static boson stars discussed in Sec.~\ref{Sec:boson stars} this quantity is given by
\begin{equation}
Q = \frac{\kappa\omega}{2}\int\limits_0^\infty |\psi|^2\frac{\gamma}{\alpha} r^2 dr.
\end{equation}
It should be interesting to analyze the implications of the conserved quantity $Q$ for the scalar field configurations surrounding a black hole constructed in this article.


\bibliographystyle{bibtex/prsty}
\bibliography{referencias.bib}
\bibliographystyle{unsrt} 

\end{document}